\documentclass[superscriptaddress,onecolumn,
amssymb,amsmath,nobibnotes,aps,prd,
nofootinbib]{revtex4}
\pdfoutput=1
\usepackage{graphicx,subfigure,bm,color,psfrag,hyperref}
\usepackage{amsfonts}
\usepackage{lipsum}
\usepackage{mathtools}
\usepackage{verbatim}
\usepackage{color}
\DeclareUnicodeCharacter{2212}{-}
\begin{document}

\title{Do current observations support transient acceleration of our universe?}

\author{Yu. L. Bolotin}
\email{ybolotin@gmail.com}
\author{V. A. Cherkaskiy}
\email{vcherkaskiy@gmail.com}
\author{M. I.  Konchatnyi}
\email{konchatnij@kipt.kharkov.ua}
\affiliation{NSC ``Kharkiv Institute of Physics and Technology'', Akademicheskaya Str. 1,	Kharkiv, 61108,	Ukraine}

\author{Supriya Pan}
\email{supriya.maths@presiuniv.ac.in}
\affiliation{Department of Mathematics, Presidency University, 86/1 College Street, Kolkata 700073, India}

\author{Weiqiang Yang}
\email{d11102004@163.com}
\affiliation{Department of Physics, Liaoning Normal University, Dalian, 116029, P. R.  China}

\begin{abstract}
In the present article we have investigated a very natural question regarding the dynamics of the universe, namely, the possibility of its decelerating phase immediately after the present accelerating phase. To begin with, we have focused on the matter creation theory which is considered to be a viable alternative to dark energy and modified gravity models. Moreover, we have introduced the cosmographic approach which allows us to express the free parameters of a cosmological model in terms of the known cosmographic parameters. Assuming a generalized matter creation rate we have discussed the theoretical bounds on the model parameters allowing the future deceleration of the universe. Moreover, using the observational bounds on the cosmographic parameters  obtained from the low redshifts observational probes, we have also examined the chance of a decelerating phase of the universe. Finally, considering a variety of known cosmological models and parametrizations, we have tested the same possibility. Our analysis shows that the chance of a future decelerating expansion of the universe is highly dependent on the choice of the cosmological models and parametrizations and also on the observational data. Even though the future decelerating expansion is allowed in some cosmological frameworks, but we do not see any strong evidence in favor of this. Perhaps, the future cosmological surveys could offer some more information regarding the fate 
of the universe. 
\end{abstract}

\maketitle

\section{Introduction}

Due to  the combined efforts of both theory and observations, our understanding on the dynamics of our universe has  changed a lot. The discovery of an accelerated expansion of the universe  \cite{Riess:1998cb,Perlmutter:1998np,Hinshaw:2012aka,Suzuki:2011hu,Planck:2018vyg} has  stimulated the scientific world, and at the same time, raised many questions that are still unknown to the community. For instance, according to the available observational evidences, the universe   should have experienced an accelerated expansion, much earlier to the present accelerating phase (late-time accelerated expansion),  known as inflation  \cite{Starobinsky:1979ty,Starobinsky:1980te,Guth:1980zm,Linde:1981mu,Mukhanov:1981xt,Barrow:1981pa,Hawking:1982cz,Starobinsky:1982ee,Guth:1982ec,Starobinsky:1983zz} in order to explain a number of cosmological puzzles related to the early-evolution of the universe, also see \cite{Barrow:1982kr,Linde:1984ir,Linde:1985ub,Kofman:1986wm,Burd:1988ss,Barrow:1990vx,Barrow:1993hn,Barrow:1995xb,Barrow:2016qkh,Paliathanasis:2017apr}. In between inflation and the current accelerating phase, the structure formation of our universe needs a cosmic phase with a decelerating expansion. 
Thus, a large amount of observational data together with the theoretical arguments greatly supported that our universe has experienced two transitions -- 
a transition from the early inflationary phase to the intermediate decelerated expansion -- and then -- a transition from the  intermediate matter dominated decelerating phase to the present accelerating phase. To describe the above trajectory of the universe a number of cosmological models have been introduced so far and they have been confronted with the observational data, see for instance \cite{Peebles:1998qn,Banerjee:2000mj,Peebles:2002gy,Pan:2013rha,Paliathanasis:2015gga,Paliathanasis:2015cza,deHaro:2016hpl,deHaro:2016cdm,Paliathanasis:2017efk,Yang:2017yme,Das:2017gjj,Yang:2018euj,Pan:2017ent,Das:2018bzx,Barrow:2016wiy,Yang:2017zjs,Haro:2019gsv,Yang:2018qec,Yang:2019jwn,Yang:2019qza,Paliathanasis:2019hbi,Das:2018ylw,Paliathanasis:2018vru,Haro:2019peq,Giacomini:2020grc, Pan:2020mst} (see the review articles in this direction \cite{Sahni:1999gb,Sahni:2006pa,Copeland:2006wr}).

Following the history of transitions of our universe, that means from  (i) its early accelerating phase to the intermediate decelerating phase and then (ii) from the intermediate decelerating phase to the presently observed accelerating phase, one natural question that one may ask is the following: do we expect another transition of our universe from its present accelerating expansion to the future decelerating phase?  This is a very crucial question in the context of modern cosmology since if our universe starts decelerating again in near future then we have to understand the immediate consequences of this decelerating phase. With the accumulation of a large number of potential cosmological probes, one may try to extract some information about the possibility of a decelerating phase of our universe in the near future. Even though the standard $\Lambda$-cold dark matter ($\Lambda$CDM) cosmology, which gives a fantastic fit to a large number of observational data, predicts
an eternal cosmic acceleration, however, we recall that the main ingredients of $\Lambda$CDM, namely dark matter and dark energy are not clearly understood yet. Additionally, with the improvement of the observational data, we have witnessed the challenges to  the $\Lambda$CDM cosmology for several observational discrepancies, see the recent reviews \cite{DiValentino:2021izs,Perivolaropoulos:2021jda}. As a consequence,  alternative cosmological models beyond $\Lambda$CDM are getting massive attention to describe the dynamical history of the universe in agreement with the latest observations.   
Among various alternatives to the standard $\Lambda$CDM cosmology, the cosmological scenario driven by the gravitationally influenced adiabatic matter creation is an excellent alternative for various reasons.  In the next paragraph we describe the potential nature of the particle creation theory.

The mechanism of matter creation process is totally different from the conventional 
dark energy (within General Relativity) and modified gravity theories $-$  the two well known approaches to understand the late-time accelerating phase of the universe \cite{Steigman:2008bc,Lima:2012cm,Lima:2014qpa,Ramos2014,Lima2016, Pan2016}. Since without any need of dark energy and gravity modifications, matter creation theory can explain the present accelerating expansion, hence, it is believed to be a third alternative to the dark energy and modified gravity theories. The basic ingredient of the matter creation theory is the creation rate of particles which modifies the gravitational field equations.  
Apart from explaining the present accelerating phase within this context \cite{Steigman:2008bc,Lima:2012cm,Lima:2014qpa,Ramos2014,Lima2016, Pan2016}, the early inflationary phase \cite{Gunzig1998,Abramo:1996ip} can also be depicted. In fact, the initial singularity 
of the universe, can also be traced out with the choice of a constant matter creation rate \cite{deHaro:2015hdp}. Interestingly, with the suitable choice of the creation rate it is possible to describe the entire evolution of the universe \cite{Lima:2012mu,Perico:2013mna}. Therefore, the theory of matter creation naturally gained enormous attention due to its potential behavior and emerged as one of the possible routes to explain the dynamical history of the universe without any need of dark energy or modified gravitational theories.

In the present article we have considered matter creation theory to investigate the possibility of future deceleration. In other words we shall test the transient nature of the cosmic acceleration. 
However, an important question that one may ask is, how the prediction of transient acceleration is reliable within the present theoretical framework? Keeping this issue in mind, we have taken a novel approach where the determination of the transient nature of the present accelerating phase arises through a model independent diagnostic -- the cosmography \cite{Weinberg1972,Visser2005,Visser2004,Dunajski2008,Guimaraes:2010mw,Bolotin2012,Bolotin2016,Dunsby2016,Bolotin2018}. In cosmography, one deals with the minimal assumption of our universe, namely the homogeneous and isotropic principle and makes use of the kinematical/geometrical quantities to understand the dynamical history of the universe in a model independent manner. The most interesting and crucial point is that, one can express the model parameters in terms of the kinematical quantities, and thus, more correct determination within the model framework is possible.   This technique has been applied in this article.

The article is organized as follows: In section \ref{sec-cosmographic-method} we describe the cosmographic parameters with a brief historical background of them. Then in section \ref{sec-constraints} we provide the observational constraints of the cosmographic parameters using the recent geometrical probes. After that in  section \ref{sec-matter-creation} we describe in detail the matter creation theory and consequently in the next section \ref{sec-model} we first introduce a very generalized matter creation model and relate the model parameters with the cosmographic parameters. We then investigate our interesting question whether the universe may decelerate in future again in section \ref{sec-future-deceleration}. Finally, we conclude the present work with a brief summary of our investigations in section \ref{sec-conclu}

\section{Background: The Cosmographic Method}
\label{sec-cosmographic-method}

The fundamental characteristics employed to trace the dynamical evolution of the universe can be either kinematical or dynamical. In the first approach, the governing quantities are directly extracted from the space-time metric while the latter approach relates with the properties of the fields filling the matter sector of the universe. Therefore, one can clearly realize that the dynamical characteristics are model dependent while the kinematic characteristics are more universal. Additionally, the kinematic quantities are free from the uncertainties compared to the physical quantities, for example, the energy densities of various fields, where such uncertainties may appear. That is why the kinematic quantities are convenient for describing the expansion history of the universe. The kinematics of cosmological expansion of a homogeneous and isotropic universe has been called cosmography \cite{Weinberg1972}. Therefore, in cosmography the minimal assumption that is need is the Friedmann-Lema\^{i}tre-Robertson-Walker (FLRW) geometry of our universe. 

In the early sixties of the last century, Allan Sandage \cite{Sandage1962} defined  the primary goal of the cosmologists is, to  search for two key parameters of the universe, namely, the Hubble parameter, $H$, and the deceleration parameter, $q$. However, an expansion with a constant acceleration is not the only possible realization of the kinematics of a non-stationary universe. A conscious mind can realize that, as the universe evolves, the relative content of the components filling the universe sector should be changing, and as a consequence, the kinematics of the expansion changes which finally results in changes in the cosmic  acceleration. Thus, for a more complete description of the kinematics of the cosmological expansion, it is useful to consider an extended set of parameters including the higher-order time derivatives the scale factor, $a(t)$. This has been investigated earlier by various researchers
\cite{Sahni:2002fz,Alam:2003sc,Visser2004,Visser2005,Arabsalmani:2011fz,Bolotin2012,Dunsby2016}. Following the earlier works, we recall the higher order derivatives of the scale factor as follows: 
\begin{align}
	\nonumber H(t)&\equiv\frac1a\frac{da}{dt};\\
	\nonumber q(t)&\equiv-\frac1a\frac{d^2a}{dt^2}\left[\frac1a\frac{da}{dt}\right]^{-2};\\
	\label{eq1_8} j(t)&\equiv\frac1a\frac{d^3a}{dt^3}\left[\frac1a\frac{da}{dt}\right]^{-3};\\
	\nonumber s(t)&\equiv\frac1a\frac{d^4a}{dt^4}\left[\frac1a\frac{da}{dt}\right]^{-4};\\
	\nonumber l(t)&\equiv\frac1a\frac{d^5a}{dt^5}\left[\frac1a\frac{da}{dt}\right]^{-5};
\end{align}
which are known as the cosmographic parameters. They all are model independent\footnote{We recall that they are called model independent under the minimal assumption of the FLRW universe. }  and apart from the Hubble parameter, remaining kinematic quantities (equivalently, cosmographic parameters) are dimensionless. The newly introduced parameters, $j$, $s$ and $l$  are respectively known as jerk parameter, snap parameter and the lerk parameter.  At this point let us clarify that in  Ref. \cite{Sahni:2002fz}, the kinetic quantities $j,\; s$ were introduced in the name of statefinders but let us note that the  parameter $s$  of eqn. (\ref{eq1_8}) was defined in a different way: $s  = (j-1)/[3 (q- 1/2)]$. 
The jerk parameter is one of the important kinematic quantities because it characterizes the evolution of the deceleration parameter. The allowance of the higher derivatives of the  scale factor in the kinematic quantities has two positive sides. Firstly, it
reflects the continuous progress of the observational cosmology and since they are model independent,  they are indeed very important in the understanding of the expansion history of the universe. Additionally,  the extension of the kinematic quantities beyond $H$ and $q$ is motivated to obtain more precise information about the dynamical history of the universe which was characterized only by $(H, q)$. 

The deceleration parameter $q$ can be related to the Hubble parameter $H$ by the following equivalent relations:
\begin{eqnarray}\label{alpha}
&&	q(t)=\frac{d}{dt}\left(\frac{1}{H}\right)-1,\\
	&& q(z)=\frac{1+z}{H}\frac{dH}{dz}-1 = \frac{1}{2}(1+z)\frac{1}{H^2}\frac{dH^2}{dz}-1
	=\frac{1}{2}\frac{d\ln H^2}{d\ln(1+z)}-1. 
\end{eqnarray}
The time derivatives of the Hubble parameter can also be expressed in terms of the cosmographic parameters:
\begin{align}
	\nonumber\dot H&=-H^2(1+q);\\
	\label{hubble_paramters_time_derivatives}\ddot H&=H^3(j+3q+2);\\
	\nonumber\dddot H&=H^4(s-4j-3q(q+4)-6);\\
	\nonumber\ddddot H&=H^5(l-5s+10(q+2)j+30(q+2)q+24).
\end{align}
Dunajski and Gibbons \cite{Dunajski2008} proposed an ingenious approach aiming to test the cosmological models satisfying the cosmological principle. The implementation of the method adopts the following sequence of steps \cite{Bolotin2016}:
\begin{enumerate}
	\item The first Friedman equation is transformed to an ordinary differential equation for the scale factor. This is achieved using the conservation equation for each component included in the model to find the dependence of the energy density on the scale factor.
	\item The resulting equation is differentiated (with respect to the cosmic time) as many times as there are free parameters in the model.
	\item The time derivatives of the scale factor are expressed through the cosmographic parameters.
	\item Solving the obtained system of linear algebraic equations, we express all free parameters of the model in terms of the cosmographic parameters.
\end{enumerate}
The procedure under consideration can be made more universal and effective if the system of Friedmann equations for the Hubble parameter $H$ and its time derivative $\dot H$ is considered as the starting point. Differentiating the equation as many times as required (this number is determined by the number of free parameters in the model), we obtain a system of equations including higher time derivatives of the Hubble parameter, such as $\ddot H$, $\dddot H$, $\ddddot H$, and so on. These derivatives are directly related to the cosmographic parameters by the relations (\ref{hubble_paramters_time_derivatives}).
The proposed approach to find the parameters of cosmological models has many advantages. Let's briefly dwell on them.
\begin{enumerate}
	\item Universality: The method is applicable to any cosmological model which satisfies the cosmological principle. This procedure can also be applied to the models allowing non-gravitational interaction beteween the fluids \cite{Bolotin2016}.
	
	\item Reliability: The results are accurate since they follow from the identity transformations.
	
	\item The procedure is very very nice and simple. 
	
	\item Free parameters of the models under consideration can be  expressed using the model indepedent cosmographic parameters. There is no need to introduce any additional parameters to perform this procedure.
	
	\item The method is very elegant in the sense that it provides an interesting way to calculate the higher cosmographic parameters using the values of lower ones. 
	
	\item The method also offers a simple test in order to analyze the compatibility of the cosmological models under consideration. Since the cosmographic parameters are model independent and hence they are universal, thus, the models are compatible only in  a non-zero intersection domain of their parameter space.
\end{enumerate}

To be more specific on the latter point, the model compatibility analysis consists of two steps. In the first step, the model parameters are expressed in terms of the cosmographic parameters. The second step is to find the intervals of cosmological parameter changes that can be realized within the framework of the considered model. Since the cosmographic parameters are universal, the models are considered compatible only in the case of a nonzero intersection of the obtained intervals.

\section{Observational data, methodology and constraints on the cosmographic parameters}
\label{sec-constraints}

In this section we describe the observational datasets that we use to constrain the cosmographic parameters. We use only the geometric datasets in this work since the datasets dependent on some fiducial cosmological model may bias the constraints. We start with the following datasets:

\begin{itemize}

    \item {\bf Hubble parameter measurements:}
    The measurements of Hubble parameter at different redshifts during the evolution of the universe play a very crucial role to understand its evolution history. In this article we use the measurements of the Hubble parameter at different redshifts in a model independent way, known as 
    the cosmic chronometer approach. The cosmic chronometers (CC) are the  most massive and passively evolving galaxies in the universe. The underlying mechanism to measure the Hubble parameter values at different redshifts is the following.  One needs to determine the quantity $dz/dt$, where $z$ stands for the redshift, and using the relation \[H(z)= -\frac1{1+z}\frac{dz}{dt},\] one can estimate the value of the Hubble parameter at some specific redshift. The estimation of the differential $dz$ is acquired through the spectroscopic method with extremely high accuracy and the differential age evolution $dt$ of such galaxies can also be accurately measured. Thus, the measurements of the Hubble parameter are considered to be model independent. We refer to  \cite{Moresco:2016mzx} for a detailed explanation of the techniques. In this work, we make use of $30$ measurements of the Hubble parameter spanned in the redshift interval $(0, 2)$  \cite{Moresco:2016mzx}. In this article we refer to this dataset as CC.
    
    \item {\bf Supernovae Type Ia:} The expansion of our universe in an accelerating manner was first indicated through the observations of  Supernovae Type Ia (SNIa). These are the geometric probes to understand the late time evolution of the universe and also known as the standard candles. There are different catalogues of SNIa. In the present article,  we use the most recent compilation of SNIa, namely the Pantheon sample. This sample comprises of 1048 data points distributed in the redshift region $z \in [0.01, 2.3]$ ~\cite{Scolnic:2017caz}. 
   
    \item {\bf $H_0$ from the Hubble Space Telescope:} We also include a gaussian prior on the Hubble constant obtained from the Hubble Space Telescope yielding $H_0 = 74.03 \pm 1.42$ km/s/Mpc at $68\%$ CL~\cite{Riess:2019cxk}. In this article we refer to this dataset as HST. 
    
    \end{itemize}
 \begingroup                                                     
\squeezetable                                               
\begin{center}                                             
\begin{table*}                                              
\begin{tabular}{cccccccccc}                                    
\hline\hline                                               
Parameters & CC & Pantheon & Pantheon+HST & Pantheon+CC & Pantheon+HST+CC \\ \hline

$H_0$ & $   69.00_{-    3.97-    7.36}^{+    3.95+    7.72}$ & $   70.03_{-   20.03-   20.03}^{+   19.972+   19.97}$ & $   72.90_{-    1.78-    3.54}^{+    1.78+    3.54}$ & $   68.49_{-    2.65-    5.29}^{+    2.61+    5.23}$ & $   71.53_{-    1.46-    2.93}^{+    1.48+    2.84}$ \\

$q_0$ & $   -0.51_{-    0.49-    0.49}^{+    0.51+    0.51}$ & $   -0.49_{-    0.25-    0.41}^{+    0.23+    0.43}$ & $   -0.49_{-    0.23-    0.41}^{+    0.23+    0.42}$ & $   -0.44_{-    0.06-    0.16}^{+    0.09+    0.15}$ & $   -0.39_{-    0.05-    0.11}^{+    0.06+    0.11}$ \\

$j_0$ & $   -0.14_{-    1.70-    3.14}^{+    1.68+    3.19}$ & $   -0.48_{-    3.50-    5.66}^{+    3.21+    5.95}$ & $   -0.44_{-    3.50-    5.54}^{+    3.0+    5.81}$ &  $   -0.77_{-    1.42-    2.36}^{+    1.01+    2.59}$ & $   -1.71_{-    0.94-    1.68}^{+    0.80+    1.74}$ \\

$s_0$ & $  -11.08_{-   14.68-   25.37}^{+   11.77+   26.70}$ & $  -12.80_{-   29.03-   36.79}^{+   13.11+   49.87}$ & $  -12.73_{-   28.23-   36.23}^{+   12.85+   48.98}$ & $  -16.14_{-   11.86-   20.11}^{+    8.49+   21.98}$ & $  -23.37_{-    8.27-   14.07}^{+    6.96+   15.62}$ \\
\hline\hline                                                                                                                    
\end{tabular}                                                                                                                   
\caption{68\% and 95\% confidence-level (CL) constraints on various cosmographic parameters, namely, $H_0$, $q_0$, $j_0$ and $s_0$, using different geometric datasets and their combinations. }
\label{tab:cosmography}                                                                                                   
\end{table*}                                                                                                                     
\end{center}                                                                                                                    
\endgroup   
\begin{figure*}
    \centering
    \includegraphics[width=0.6\textwidth]{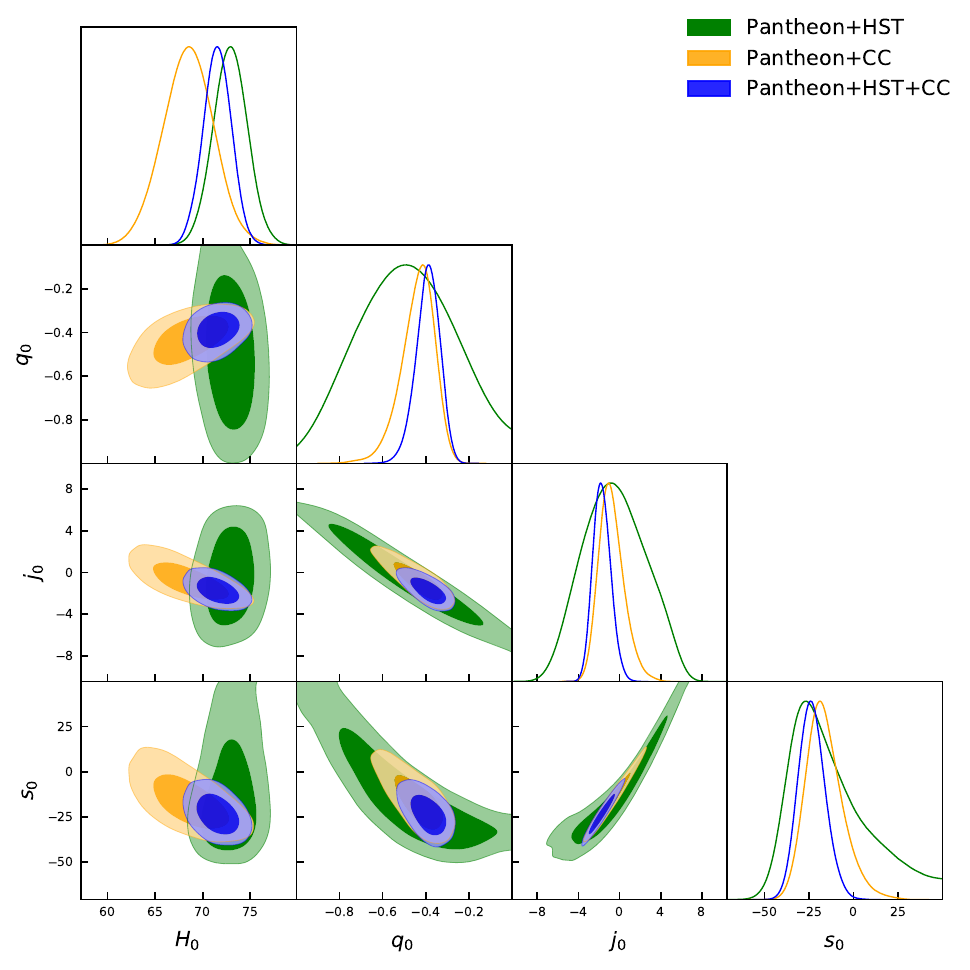}
    \caption{One-dimensional marginalized distributions of the individual cosmographic parameters and the two-dimensional joint contours at 68\% and 95\% CL for the combined datasets Pantheon+HST, Pantheon+CC and Pantheon+HST+CC. }
    \label{fig:contour}
\end{figure*}
\begin{figure*}
    \includegraphics[width=0.45\textwidth]{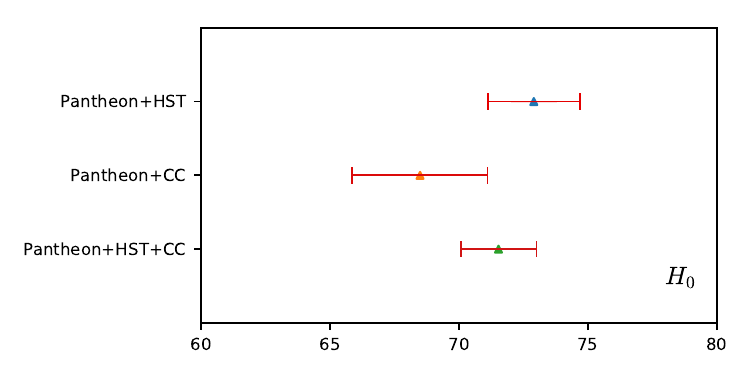}
    \includegraphics[width=0.45\textwidth]{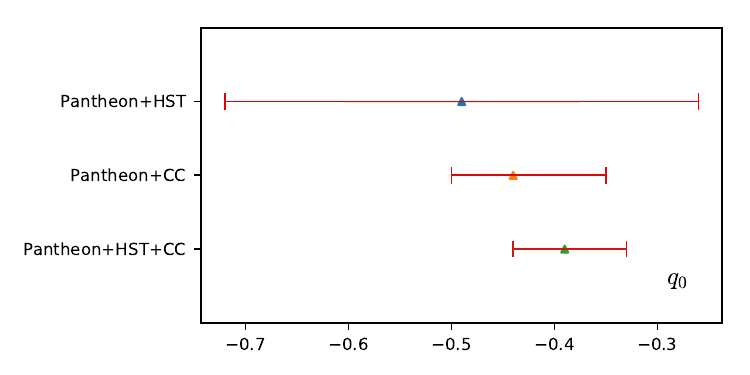}
    \includegraphics[width=0.45\textwidth]{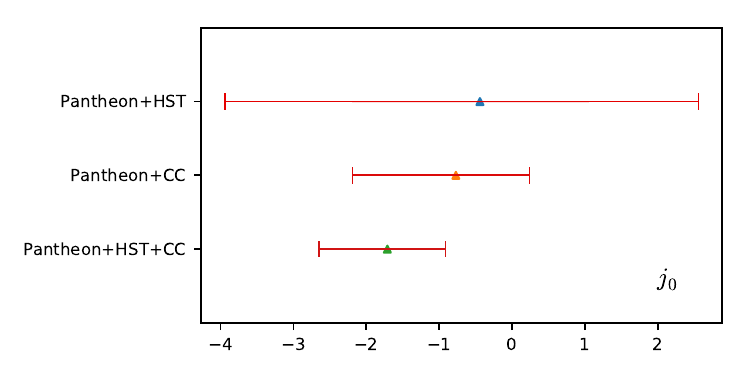}
    \includegraphics[width=0.45\textwidth]{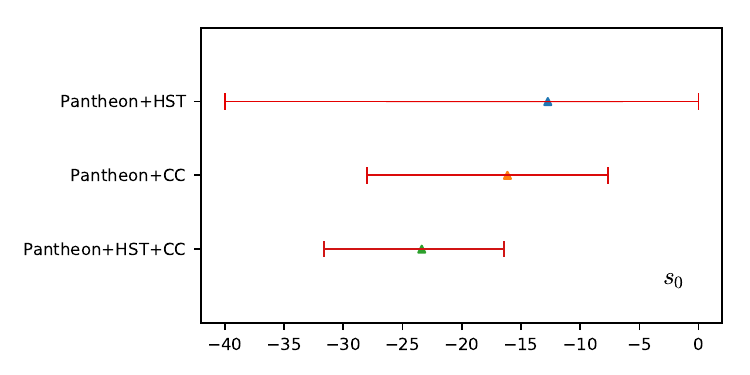}
    \caption{Whisker graphs at 68\% CL representing the cosmographic parameters $H_0$, $q_0$, $j_0$, $s_0$  for the combined datasets Pantheon+HST, Pantheon+CC and Pantheon+HST+CC as given  in Table \ref{tab:cosmography}. The upper left graph and upper right graph respectively stands for the parameters $H_0$ and $q_0$ whilst the lower left graph and lower right graph respectively displays the parameters $j_0$ and $s_0$. }
    \label{fig:whisker}
\end{figure*}
  
With the above datasets at hand we now proceed to describe the underlying mechanism of constraining the cosmographic parameters.
It is well known that for a spatially flat FLRW universe, the luminosity distance can  be expanded using the redshift $z$ together with the cosmographic
parameters as follows
\begin{widetext}
\begin{eqnarray}\label{series1}
d_L(z) &=& c H_0^{-1}\left\{z+(1-q_0)z^2/2-\left(1-q_0-3q_0^2+j_0\right)z^3/6\right.\nonumber \\ && \nonumber \\
&+&\left.\left[2-2q_0-15q_0^2-15q_0^3+5j_0+10q_0j_0+
s_0\right]z^4/24+...\right\}.
\end{eqnarray}
\end{widetext}
where $c$ is the velocity of light and the sub-index $0$ attached to any quantity refers to its present value.  One can clearly see that the above infinite series may diverge for  $z > 1$.  In order to avoid with such divergence issues related to the highest
redshift objects, one can introduce a new variable $y=z/(1+z)$, and consequently recast the above series (\ref{series1}) into the following form \cite{Cattoen:2007sk}
\begin{widetext}
\begin{eqnarray}\label{series2}
d_L(y)&=&cH_0^{-1}\left\{y+(3-q_0)y^2/2+(11-j_0-5q_0+3q_0^2)y^3/6\right.\nonumber\\
&+&\left.\left(50-7j_0-26q_0+10q_0j_0+21q_0^2-15q_0^3+s_0\right)y^4/24+\mathcal{O}(y^5)\right\}.
\end{eqnarray}
\end{widetext}
Thus, with the change of variable $z :\rightarrow y$, the previous interval $z\in (0,\infty)$ is now mapped into
$y\in(0,1)$. Now, we move toward the observational constraints on the cosmographic parameters.

In order to constrain the key parameters of this work, we have used a very efficient cosmological package   \texttt{CosmoMC}~\cite{Lewis:2002ah,Lewis:1999bs}. This is a Markov chain Monte Carlo package having a convergence diagnostic by Gelman and Rubin \cite{Gelman-Rubin}. Let us now summarize the observational constraints on the cosmographic parameters using the cosmological datasets shown above.

In Table~\ref{tab:cosmography} we have shown the constraints on ($H_0, q_0, j_0, s_0$) considering various observational datasets, namely, CC, Pantheon, Pantheon+HST, Pantheon+CC and Pantheon+HST+CC. From Table~\ref{tab:cosmography}, one can clearly notice that combined datasets offer more constraining power compared to CC and Pantheon alone. Thus, from now on we shall discuss the results obtained from the combined datasets which are summarized in the last three columns of Table~\ref{tab:cosmography}. Following this, 
in Fig. \ref{fig:contour} we have displayed the one-dimensional marginalized posterior distributions of the cosmographic parameters and the two-dimensional joint contours only for the combined datasets Pantheon+HST, Pantheon+CC and 
Pantheon+HST+CC. From Table~\ref{tab:cosmography}, we can see that for Pantheon+HST, we have a large value of $H_0 = 72.90_{- 1.78}^{+ 1.78}$ (at 68\% CL) which is mainly influenced by the gaussian prior on $H_0$ from HST, together with 
$q_0 = -0.49_{-    0.23}^{+    0.23}$ (68\% CL) and $j_0 = -0.44_{-    3.50}^{+    3.0}$ (68\% CL). However, for Pantheon+CC, we see that although the Hubble constant value is slightly shifted towards the lower value ($H_0 = 68.49_{-2.65}^{+    2.61}$ at 68\% CL for Pantheon+CC) compared to its estimation from Pantheon+HST, but it is still higher compared to the estimation by the Planck team (within the $\Lambda$CDM paradigm) \cite{Planck:2018vyg}. On the other hand, the parameters $q_0$, $j_0$, $s_0$ are significantly improved due to the reduction of the error bars compared to Pantheon+HST analysis. This effect is clearly seen from Fig.~\ref{fig:contour} if we compare the one-dimensional posterior distributions of $q_0$, $j_0$, $s_0$ for Pantheon+CC and Pantheon+HST. Additionally, from Fig.~\ref{fig:contour}, one can further notice that the inclusion of CC makes $q_0$, $j_0$, $s_0$ correlated with $H_0$ and this correlation was absent in Pantheon+HST (see Fig.~\ref{fig:contour}). For the final combined dataset, namely, Pantheon+HST+CC, we again notice that the error bars on $q_0$, $j_0$, $s_0$ are again improved compared to Pantheon+CC. As $H_0$ is correlated with the remaining cosmographic parameters, namely, $q_0$, $j_0$ and $s_0$, therefore, due to a large value of $H_0$ which is obtained in this case influenced by a gaussian prior on $H_0$ from HST, the cosmographic parameters will be affected due to the nature of the correlation with $H_0$. As $j_0$ and $H_0$ are anti-correlated with each other (see Fig. ~\ref{fig:contour}), therefore, due to a large value of $H_0$,  
$j_0$ goes down yielding $j_0 = -1.71_{- 0.94}^{+    0.80}$ (at 68\% CL). In order to be more transparent on the improvement in the parameter space, in Fig. \ref{fig:whisker} we have shown the whisker graphs of the cosmographic parameters at 68\% CL. The upper left graph and upper right graph of Fig. \ref{fig:whisker} respectively stands for $H_0$ and $q_0$ parameters while the lower left graph and lower right graph of Fig. \ref{fig:whisker} respectively stands for the $j_0$ and $s_0$ parameters. It must be mentioned that the constraints obtained in Table~\ref{tab:cosmography} are deviating from the $\Lambda$CDM model (corresponding to $j_0 =1$), however, such deviation actually depends on the observational datasets (see Ref. \cite{2011PhLB..702..114X}) as well as  on the number of terms present in the Taylor series expansion in eqn. (\ref{series2}), see for instance Ref. \cite{Li:2019qic}. Therefore, the sensitivity of the observational data and the higher order terms present in the Taylor series expansion of the luminosity distance  are the two key factors in the constraints on the cosmographic parameters.

\section{Matter creation theory: a brief overview}
\label{sec-matter-creation}

There are many ways to explain the accelerating expansion of the universe. Two well known approaches are the introduction of dark energy (within General Relativity) and the modifications of the General Relativity. The common feature in both the above approaches is the violation of strong energy condition which happens due to the presence of some component with negative pressure. Thus, the key ingredient of the accelerating expansion is the negative pressure, and this negative pressure arises  naturally  when the system digresses from the thermodynamic equilibrium. This concept was used to propose an alternative mechanism of dark energy and modified gravity theories following Zeldovich \cite{Zeldovich1970} where he suggested that negative pressure maybe generated in the particle creation process due to the energy of the gravitational field. 
This idea got significant attention in the community and inspired many researchers. In Ref. \cite{Prigogine:1989wc} the authors established that if we consider our universe as an open thermodynamic system characterized by a fluid with a non-conserved number of particles $N(t)$, then the conservation of the fluid can be described as follows
\begin{equation}
\label{eq3_41} \dot n+\Theta n=n\Gamma,
\end{equation}
where $n\equiv N (t)/V$ denotes the number density of the particles within the comoving volume $V$; $\Theta = u_{;\mu}^\mu$ refers to the velocity of the expansion of the fluid ($\Theta=u_{;\mu}^\mu=3H$ for the FLRW metric); the quantity $\Gamma$ denotes the particle creation rate in the comoving volume and this is the source of the negative pressure where $\Gamma>0$ represents the particle creation process while $\Gamma<0$ denotes the annihilation process. Naturally, $\Gamma  = 0$ is the no particle creation process.

Let us consider the spatially flat FLRW model of the universe as an open thermodynamical system, the energy momentum tensor of the fluid endowed with particle creation process can be identified as  
\begin{equation}
\label{eq3_44} T_{\mu\nu}=\left(\rho+p+ P_c \right)u_\mu u_\nu+\left(p+ P_c\right)g_{\mu\nu},
\end{equation}
where $P_c$ denotes the pressure term which arises due to production of particle at the expense of the energy of the gravitational field.

Now, one can quickly write down the Friedmann equations as follows (in the units where $8 \pi G = 1$)
\begin{align}
	H^2  = \frac{\rho}{3}, \quad \quad \dot H =-\frac12\left(\rho+p+ P_c \right). \label{eq3_45} 
\end{align}
One can also write down conservation equation
\begin{equation}
\label{eq3_46} \dot\rho+3H\left(\rho+p+P_c\right) = 0.
\end{equation}
We now restrict the creation process to be adiabatic. For adiabatic process, it is already well known that the creation pressure is related to the particle creation rate as  
\cite{Steigman:2008bc,Lima:2012cm,Lima:2014qpa,Ramos2014,Lima2016, Pan2016,deHaro:2015hdp}:

\begin{equation}
\label{eq3_49} P_c=-\frac{\Gamma}{3H}(\rho+p).
\end{equation}
This shows that for an adiabatic process, the creation pressure can be expressed in terms of the particle creation rate. That means if the particle create rate is known, one can clearly understand the nature of the creation pressure.

The effective equation of state parameter for the cosmological model with the creation of particles reads
\begin{align}
	w_{\rm eff}&=\frac{p_{tot}}{\rho_{tot}}=\frac{p+P_c}{\rho} = -1+\gamma\left(1-\frac\Gamma{3H}\right), \label{eq3_50} 
\end{align}
where we have considered the barotropic equation-of-state, $p = (\gamma -1) \rho$.  
From the above equation, one can notice that for $\Gamma<3H$,  $w_{\rm eff}>-1$, which corresponds to the quintessence fluid, and for $\Gamma>3H$, we realize the phantom dark energy ($w_{\rm eff}<-1$). The case $\Gamma=3H$ introduces the cosmological constant ($w_{\rm eff} = -1 $).

Now let us focus on one of the main parameters, namely, the deceleration parameter, using the definition of the deceleration parameter $q=-1-\frac{\dot H}{H^2}$ together with the following equation
\begin{align} 
\label{eq3_51b} 
\dot H&=-\frac32\gamma H^2\left(1-\frac{\Gamma}{3H}\right), 
\end{align}
one can express the deceleration parameter involving the creation rate $\Gamma$ as 
\begin{equation}
\label{eq3_52} q= -1+\frac32\gamma\left(1-\frac{\Gamma}{3H}\right).
\end{equation}
Now, using the Friedmann equations in (\ref{eq3_45}), one can also have an equation for the scale factor 
\begin{equation}
\label{eq3_53} \frac{\ddot a}{a}+\frac{H^2}{2}\left[1+3w-\frac{1+w}{H}\Gamma\right]=0.
\end{equation}
where we have considered $w = \gamma-1$. Note that for $\Gamma=0$, the standard second Friedmann equation can be reproduced.

The cosmological scenarios with matter creation ($\Gamma\ne0$) have been found to explain the evolutionary history of the universe including the present accelerated expansion \cite{Lima2007,Pan2013,Pan2016,Chakraborty2014,Lima2016,Ramos2014,Chakraborty14}. Moreover, within the context of matter creation models, it has been  argued that our universe may decelerate in future \cite{Chakraborty2014,Pan:2014lua}. Interestingly, some earlier data driven investigations in Ref. \cite{Guimaraes:2010mw} suggest that the possibility  of the future deceleration of the universe cannot be totally ruled out.

Now, in order to explain the dynamical  evolution of the universe within the context of matter creation theory, one needs to consider the rate of matter creation. There are several debates about the actual matter creation rate\footnote{However, the readers might be interested in Ref. \cite{Zeldovich:1971mw} where an estimation of the matter creation is given by Zeldovich and Starobinsky.}, and that is why, usually, one considers some specific phenomenological choices for $\Gamma (H)$ and fits the model with the observational datasets. Here, we consider a very general choice for $\Gamma$, that depends on the Hubble rate, $H$.

We propose a very general matter creation model which includes a number of matter creation models as follows \cite{Chakraborty14}: 
\begin{eqnarray}\label{Gamma-power-series}
\Gamma(H)=\sum\limits_i\Gamma_iH^i,
\end{eqnarray}
where $\Gamma_i$'s are constants and the index $i$ could take positive, negative values including zero. So, naturally, with the choice of different values of $i$, one can generate a number of matter creation models. The model $\Gamma (H) = \Gamma_0$ is of special importance because the particle creation rate, $\Gamma (H)$, does not depend on the external parameters, such as the Hubble expansion rate of the universe and others, but this particle creation rate directly depends on the intrinsic nature of the matter component. Although the current content deals with the particle creation process but this can be generalized in other cosmological theories. In Ref. \cite{Shafieloo:2016bpk} the authors first introduced the idea of metastable dark energy models where the decay rate of dark energy depends only on the intrinsic nature of dark energy and the constant decay rate was studied in the light of recent observational data.  

Having the generalized choice of the matter creation rate given above, a natural question that immediately arises is the following: How sufficiently can these models describe the dynamical history of the universe? This is a very important question because, the model 
in eqn. (\ref{Gamma-power-series}), contains a number of free parameters $\Gamma_i$ and they can influence the dynamics.  Therefore, one needs to determine the parameters. Now, the question arises, whether we need to employ some additional cosmological measurements to determine $\Gamma_i$'s or one could express them through already known quantities?  Below we will show that the free parameters $\Gamma_i$'s determining the particle creation rate can indeed be expressed analytically using the cosmographic parameters using the approach proposed in Ref.  \cite{Bolotin2018}. Hence, these unknown quantities,  namely, $\Gamma_i$'s can be estimated through the estimated values of the cosmographic parameters using the observational data. Now, recasting eqn. (\ref{eq3_51b}) in the following way
\begin{equation}
\label{eq3_54} \Gamma=3 H \left(1+\frac{2}{3(1+w)}\frac{\dot H}{H^2}\right),
\end{equation}
one could understand that  if the expansion history of the universe is prescribed in terms of the Hubble function $H (z)$, one could clearly understand how the matter creation rate evolves with the evolution of the universe. In this case,  the Hubble function function $H(z)$ can be known for some specific models, for example,  in the Standard Cosmological Model (SCM), there is no particle production process meaning that $\Gamma = 0$.  Now, a given expansion history will lead to some  ad hoc specification of the matter creation rate $\Gamma(H)$ and this is not much appealing. The simplest and an effective approach is to consider a  phenomenological choice  of the function $\Gamma(H)$, and then one can constrain this choice using the available information from the known stages of the universe's evolution.
 
In order to proceed, we break the dynamical history of the universe into three different stages: (i) the early radiation dominated universe, (ii) intermediate matter dominated (decelerating) phase and (ii) the current accelerating expansion. In Ref.  \cite{Gunzig1998}, the authors formulated a number of conditions the matter creation rate $\Gamma(H)$ should satisfy so that there is no such conflicts associated with the early evolution of the universe and $\Gamma(H)\propto H^2$ was found to be the best choice. For the intermediate matter dominated era, $\Gamma(H)\propto H$ works fine. In fact, with this choice of $\Gamma$, eqn. (\ref{eq3_51b}) can be  solved easily leading to $H\propto t^{-1}$ and consequently, scale factor has a power-law dependence with the cosmic time. With the evolution of the universe, the mechanism of matter creation should start so that the creation pressure could generate the accelerated expansion of the universe. According to the available historical records, present accelerating phase of the universe is well described by the SCM.  The Hubble function in this case satisfies the evolution law 
\begin{equation}
\label{eq3_55} \dot H+\frac32H^2\left[1-\left(\frac{H_f}{H}\right)^2\right]=0,
\end{equation}
where $H_f=\sqrt{\Lambda/3}$ being the de Sitter asymptotic value of the Hubble parameter ($H\ge H_f$) in which  $\Lambda > 0$ refers to the cosmological constant. Therefore, the explicit form of the matter creation function $\Gamma(H)$ can be found using the fact  that the evolution equation of the matter creation model characterized by the modified eqn. (\ref{eq3_51b}) leads to the cosmological evolution which coincides with  the SCM. Thus, comparing eqns. (\ref{eq3_51b}) and (\ref{eq3_55}), one finds that 
\begin{equation}
\label{eq3_56} \frac{\Gamma}{3H}=\left(\frac{H_f}{H}\right),
\end{equation}
which gives $\Gamma\propto H^{-1}$. The matter creation rate $\Gamma=3H_f$ corresponds to the de Sitter phase: $\dot H = 0$, $H = H_f = constant$. Finally, for the constant matter creation rate $\Gamma = constant$,  one obtains the initial singularity of the universe \cite{deHaro:2015hdp}. Therefore, it is evident that different functional form of the matter creation function may lead to different phases of our universe and hence they deserve investigations. Thus, considering various linear
combinations  of the terms, such as, $\Gamma= constant$, $\Gamma\propto H$, $\Gamma\propto H^2$, and $\Gamma\propto H^{-1}$, one may certainly construct a wide class of eventual  cosmological models in order to investigate the dynamical history of our universe.

\section{Cosmography of the creation model}
\label{sec-model}

In this section we explicitly show how one could determine the free parameters of any matter creation rate  function $\Gamma(H)$. We begin with the following case where the matter creation rate is inversely proportional to the Hubble function: 
\begin{equation}
\label{eq3_57} \Gamma(H)=\Gamma_0+\frac{\Gamma_{-1}}{H},
\end{equation}
where the model parameters $\Gamma_0$ and $\Gamma_{-1}$ are real constants, and as we shall show below these free parameters can be expressed in terms of the present values of the cosmographic parameters.

The system of equations needed to find the free parameters is \cite{Bolotin2018}
\begin{align}
	\nonumber \frac{\dot H}{\alpha H^2}&=1-\frac13\frac{\Gamma_0}{H}-\frac13\frac{\Gamma_{-1}}{H^2},\\
	\label{eq3_58} \frac{\ddot H}{\alpha\dot H H}&=2-\frac13\frac{\Gamma_0}{H},\quad \alpha\equiv-\frac32(1+w).
\end{align}
Now, the cosmographic parameters are related to the time derivaties of the Hubble parameter as follows
\begin{align}
	\nonumber \frac{\dot H}{\alpha H^2}&=\frac1\alpha(1+q),\\
	\label{eq3_59} \frac{\ddot H}{\alpha\dot H H}&\equiv k_1=\frac23\frac{(j+3q+2)}{(1+w)(1+q)}.
\end{align}
Hence, one can now expres the free parameters, $\Gamma_0$ and $\Gamma_{-1}$ by solving  the system of equations (\ref{eq3_58}) in terms of the cosmographic parameters as 
\begin{align}
	\nonumber \frac{\Gamma_0}{H}&=3(2-k_1)=2\frac{-j+3w(1+q)+1}{(1+q)(1+w)},\\
	\label{eq3_60} \frac{\Gamma_{-1}}{H^2}&=3\left(k_1-1-\frac23\frac{1+q}{1+w}\right)\\
	\nonumber&=-\frac{-2j+2q^2+3(1+q)+q+1}{(1+w)(1+q)}.
\end{align}
A remark is now in order: The free parameters $\Gamma_i$'s are all real constants. They can depend only on the combinations of the cosmographic parameters $H, q, j, s, \ldots$ so that $\Gamma_i$'s to remain indepdent of the time and the parameters $\Gamma_i$'s in eqn. (\ref{eq3_60}) can be calculated at any time.
This is the strength of our approach.  Obviously, the cosmographic parameters $H, q, j, s, \ldots$ should be known at this point of time. Therefore, we choose the set of cosmographic parameters 
$H_0, q_0, j_0, s_0, \ldots$ It is easy to verify that, \[\frac{d\Gamma_i}{dt}\equiv0.\] Hence, the right-hand
sides of the relations in (\ref{eq3_60}) can be calculated taking the present  values of
the cosmographic parameters, namely, $H_0, q_0, j_0, s_0, \ldots$.

We stress that the solution (\ref{eq3_60})  is exact. Nevertheless, let us test this in several ways. Recall that the deceleration parameter $q$ in this case takes the form
\begin{equation}
\label{eq3_61} q = -1+\frac32(1+w)\left(1-\frac{\Gamma}{3H}\right),
\end{equation}
which allows us to express the matter creation rate as follows:
\begin{equation}
\label{eq3_62} \Gamma=H \left(\frac{1-2q+3w}{1+w}\right).
\end{equation}
Let us note that the solutions in  (\ref{eq3_60}) are valid for any instance of time. Therefore, substituting (\ref{eq3_60}) into (\ref{eq3_57}), we must reproduce the eqn.  (\ref{eq3_62}). 

Let us consider another approach to verify the solution (\ref{eq3_60}), namely for the SCM. 
\[\Gamma\equiv0\Rightarrow\Gamma_0=\Gamma_{-1}=0, j=1, w=0.\]
It is easy to see that for $\Gamma_0 =  0$, and $\Gamma_{-1}=0$ one obtains $q=1/2$. This is the correct value of $q$ for a universe filled with cold matter.

We now consider another matter creation model in which the matter creation rate depends linearly on the Hubble rate:
\begin{equation}
\label{eq3_63} \Gamma=\Gamma_0+\Gamma_1H.
\end{equation}
Similar to the earlier model, this model also contains two free parameters, $\Gamma_0$ and $\Gamma_1$.  So, our treatment will be similar as performed above.  Therefore, following the previous approach one can find the free parameters in terms of the cosmographic parameters as 
\begin{align}
	\nonumber \frac{\Gamma_0}{H}&=2\frac{j-q(1+2q)}{(1+q)(1+w)},\\
	\label{eq3_64} \Gamma_1&=\frac{-2j+2q^2+3w(1+q)+q+1}{(1+w)(1+q)}.
\end{align}
One can easily verify that the above values of the parameters satisfy the relation (\ref{eq3_62}). Moreover, for the SCM, one has
\[\Gamma\equiv0\Rightarrow\Gamma_0=\Gamma_1=0\Rightarrow2q^2+q-1=0\Rightarrow q=\frac12,\]
which is again the correct value of $q$ for a universe filled with cold matter.

Further, we consider the following matter creation model having quadratic dependence on the Hubble rate:
\begin{equation}
\label{eq3_65} \Gamma=\Gamma_0+\Gamma_2H^2.
\end{equation}
The free parameters, $\Gamma_0$ and $\Gamma_2$, following the earlier techniques, can be expressed as 
\begin{align}
	\nonumber \frac{\Gamma_0}{H}&=\frac{2j+3q(-2q+w-1)+3w+1}{2(1+q)(1+w)},\\
	\label{eq3_66} \Gamma_2H&=\frac{-2j+2q^2+3w(1+q)+q+1}{2(1+w)(1+q)}.
\end{align}
And similarly, the test relation (\ref{eq3_62}) is valid for the above values and for the SCM we find: 
\[\Gamma\equiv0\Rightarrow\Gamma_0=\Gamma_2=0\Rightarrow2q^2+q-1=0\Rightarrow q=\frac12.\]

Finally, we consider a general matter creation model in the following way: 
\begin{equation}
\label{eq3_67} \Gamma=\Gamma_0+\Gamma_1H+\Gamma_{-1}/H.
\end{equation}
In order to express the free parameters of this model we need to include the third-order time derivative of the Hubble parameter, i.e. $\dddot H$. In this case one can find the following system of equations
\begin{align}
	\nonumber \frac{\dot H}{\alpha H^2}&=1-\frac13\left(\frac{\Gamma_0}{H}+\Gamma_1+\frac{\Gamma_{-1}}{H^2}\right),\\
	\label{eq3_68} \frac{\ddot H}{\alpha\dot H H}&=2-\frac23\Gamma_1-\frac13\frac{\Gamma_0}{H},\\
	\nonumber \frac{\dddot H}{\alpha\ddot H H}&=\left(2-\frac23\Gamma_1\right)\frac{\dot H^2}{\ddot H H}+\left(2-\frac13\frac{\Gamma_0}{H}-\frac23\Gamma_1\right).
\end{align}
Since the quantities  \(\dot H/(H^2)\), \(\ddot H/(\dot H H)\), \(\dddot H/(\ddot H H)\), and \(\dot H^2/(\ddot H H)\), in the system of equations (\ref{eq3_68}) can be expressed in terms of the dimensionless  cosmographic parameters, hence, one can now find the free parameters of the matter creation model in terms of the cosmographic parameters as 
\begin{align}
	\nonumber \frac{\Gamma_0}{H}&=-\frac{2\left[j^2+j(q(q+4)+1)+q(2q+s+1)+s\right]}{(1+q)^3(1+w)},\\
	\label{eq3_69} \Gamma_1&=\frac{j^2+2jq+3q^2+qs+3w(1+q)^3+3q+s+1}{(1+q)^3(1+w)},\\
	\nonumber \frac{\Gamma_{-1}}{H^2}&=\frac{j^2+2j(q(q+3)+1)+q(s-q(q+2)(2q+1))+s}{(1+q)^3(1+w)}.
\end{align}
One can verify the above relations following the same technique we used earlier. 
Now, considering again the SCM, one can see that the first Friedman equation for this model can alternatively be expressed as 
\[s+2(q+j)+qj=0.\]
Similarly, since  SCM ($\Lambda=0$, $\Omega_{mat}=1$) corresponds to $\Gamma_0=\Gamma_1=\Gamma_{-1}=0$, thus, one could easily verify that  $q=1/2$, $j=1$, $w=0$ which also leads to $s=-7/2$, as expected in the SCM.

\begin{figure}
\includegraphics[width=0.6\textwidth]{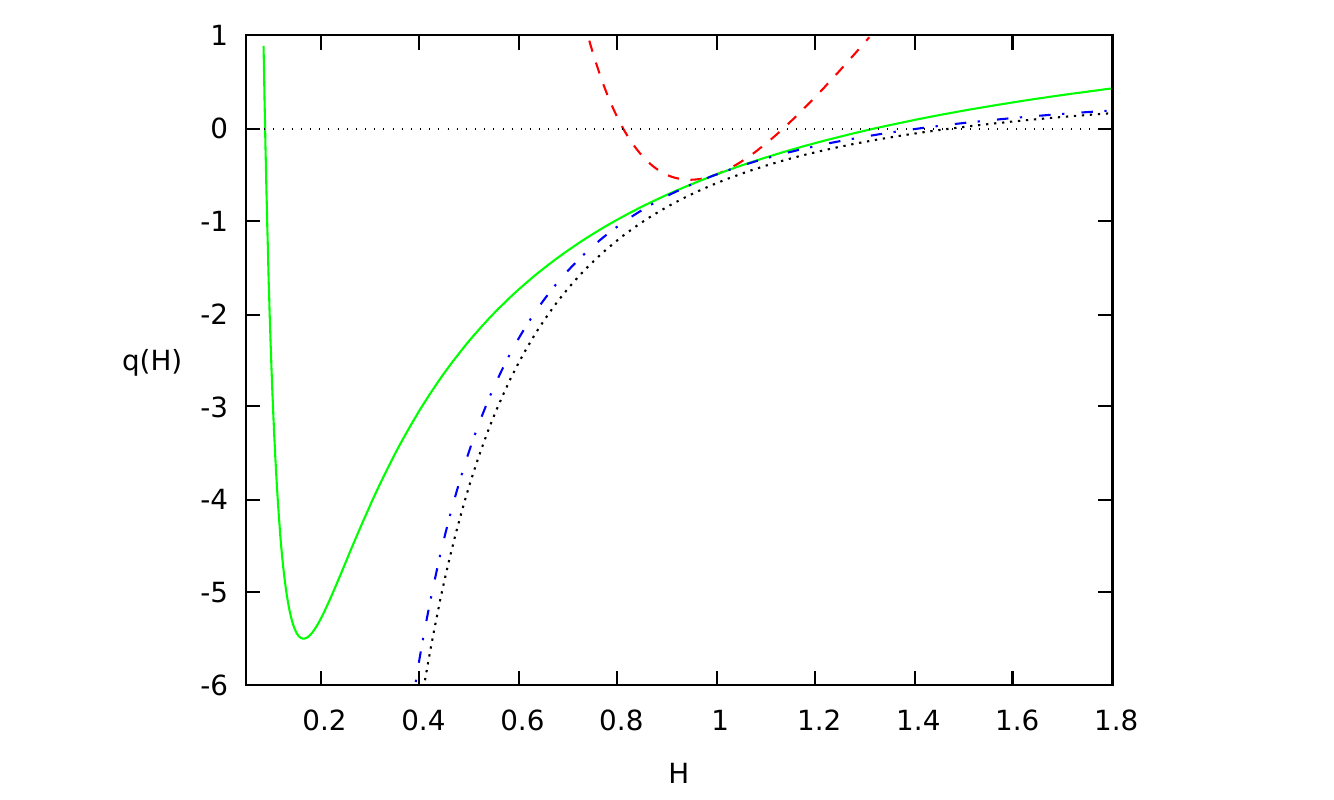}
\caption{The evolution of the deceleration parameter $q(H)$ (eqn. \ref{deceleration}) driven by the matter creation model (eqn. (\ref{eq3_67})) corresponding to the cosmographic parameter values $q_0=-0.5$, $j_0=1$, and different snap values: $s_0=-10$ (red dashed curve), $s_0=-1.1$ (green solid curve), $s_0=-0.5$ (blue dash dot curve) has been shown for a fixed value of $\gamma = 1.3$. We also compare the results with the SCM result (black dotted curve).  }
\label{Fig-deceleration1}
\end{figure}
\begin{figure}
	\includegraphics[width=0.6\textwidth]{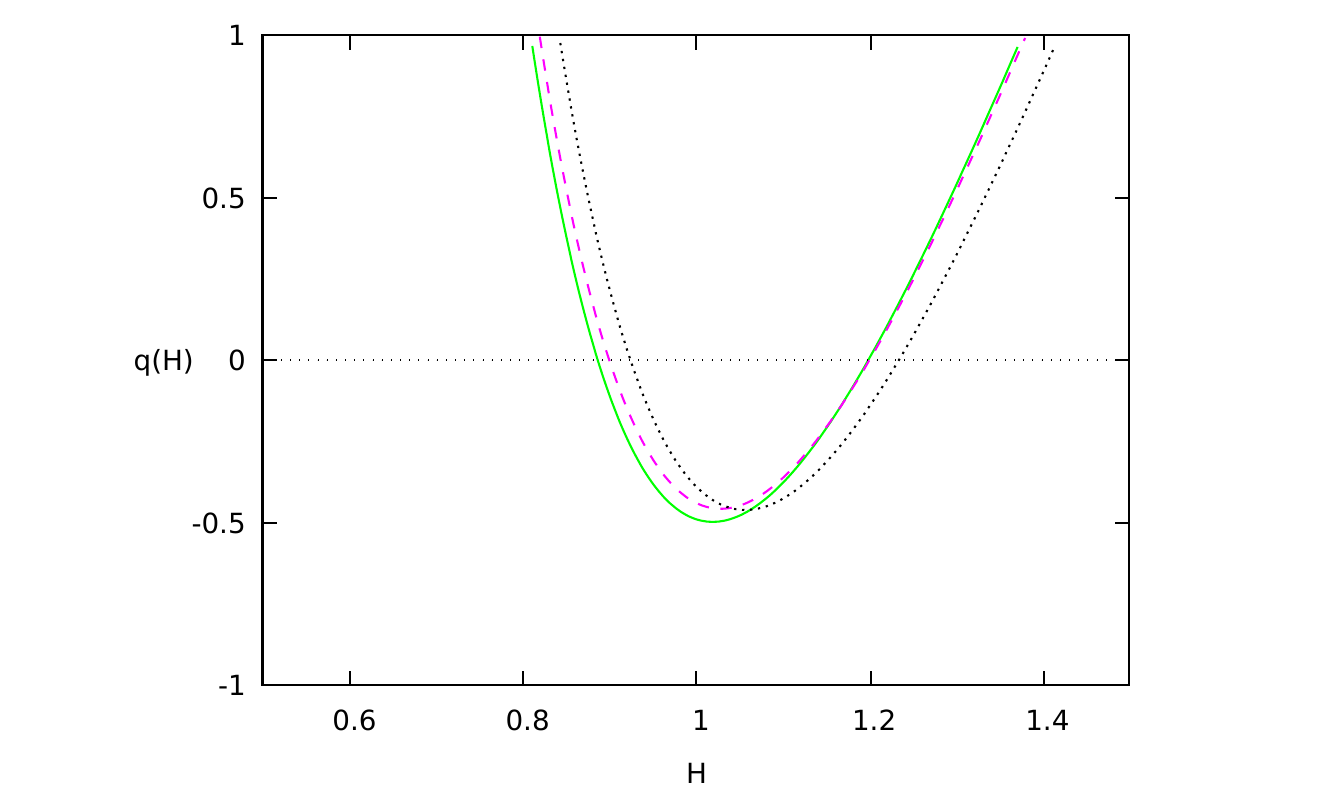}
	\caption{We show the evolution of the deceleration parameter, $q(H)$ (eqn. \ref{deceleration}) driven by the matter creation model (see eqn. (\ref{eq3_67})) using the kinematical constraints ($q_0$, $j_0$, $s_0$) obtained from the cosmological datasets, namely, Pantheon+HST (green solid curve), Pantheon+CC (magenta dashed curve) and Pantheon+HST+CC (black dotted curve) (see Table~\ref{tab:cosmography}). Let us note that the curves for Pantheon and Pantheon+HST are very close to each other and hence they are hardly distinguished from one another. We mention that similar to the earlier Fig. \ref{Fig-deceleration1}, we have fixed $\gamma = 1.3$ for drawing the curves.  }
		\label{Fig-deceleration2}
\end{figure}
\begin{figure*}
	\includegraphics[width=0.6\textwidth]{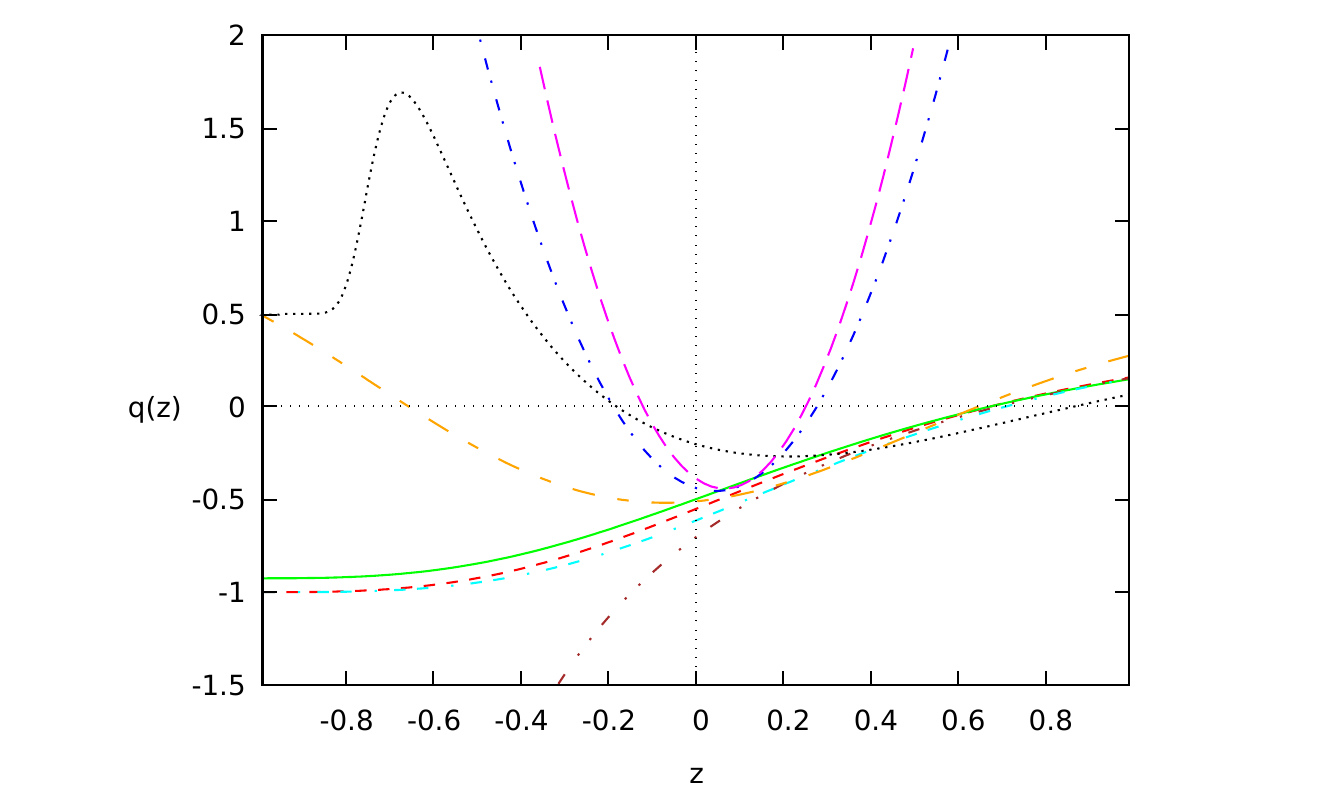}
	\caption{We show the evolution of the deceleration parameter for a variety of cosmological models, namely, $\Lambda$CDM (red dashed curve), $w$CDM  ($w = −0.95$, green solid curve), $w_0w_a$CDM ($w_0= -0.67$, $w_a = - 1.05$, black dotted curve)  together with three phenomenological parametrizations of $q(z)$ including $q_{I} (z)$  for $q_0 = −0.701$, $q_1 = 1.718$ (brown dash double-dot curve), $q_{II} (z)$ for $q_1 = 2.87$, $q_2 = 3.27$ (cyan double-dash dot curve), $q_{III} (z)$ for $q_0 = −0.51$, $z_c = 0.61$ (orange dash long-dash curve) as well as cosmographic expression of the deceleration parameter in eqn. (\ref{q-model-independent}). For the cosmographic expression of the deceleration parameter (i.e. eqn. (\ref{q-model-independent})), we have considered the values of ($q_0, j_0, s_0$) from Pantheon+CC (blue dash dot curve) and Pantheon+HST+CC (magenta long dashed curve), see Table~\ref{tab:cosmography}.  }
		\label{Fig-deceleration3}
\end{figure*}
\section{Will there be any deceleration of our Universe in near future?}
\label{sec-future-deceleration}

The dynamics of the universe is not clearly understood yet. The possibility of future deceleration of our universe is itself a very interesting question because if our universe really decelerates in near future, the fate of the universe will be different compared to an eternal accelerating phase of the universe as determined within the $\Lambda$CDM cosmological paradigm. Since $\Lambda$CDM is not the final picture of our universe which is clear from the challenges to $\Lambda$CDM that we are witnessing over the years, hence,
in the process of searching for alternative cosmological models beyond $\Lambda$CDM, everything is possible. Perhaps our universe may decelerate in near future or it could accelerate forever. 
With the growing sensitivity of the observational probes, one can try to understand whether the available observational data indicate any sign of future deceleration of the universe or not. We recall that a set of observational data has indicated the slowing down of the cosmic acceleration \cite{Starobinsky2009,Cardenas:2013roa,Magana:2014voa,Shahalam:2015lra,Hu:2015ksa}. Even though the model-independent reconstruction of the expansion history of the universe says differently about the slowing down of the cosmic acceleration \cite{Zhang:2016tto,Zhang:2017jvo}, however, as argued by the authors, the nature of the cosmic acceleration can be determined with more accuracy through the inclusion of cosmic microwave background radiation, baryon acoustic oscillations and other datasets. This is indeed a valid statement by the authors because it is not possible to conclude the nature of the cosmic acceleration from the use of a few number of observational data.
Thus, this question really demands a thorough investigations at this stage from various directions. In this section we try to examine this issue in light of the theoretical developments in cosmology as well as from the observational data.  As the theory of matter creation is  one of the viable cosmological scenarios having the potentiality to describe different phases of the universe with a suitable choice of the matter creation rate, therefore,  we start our discussions taking this particular theory and then consider other cosmological scenarios.  
We begin with the evolution of the deceleration parameter, $q$, which in the present framework, that means in presence of the matter creation, can be written as the function of the Hubble parameter $H$:
\begin{eqnarray}
q\equiv-\left(1+\frac{\dot H}{H^2}\right)=-1+\frac{3 \gamma}2\left(1-\frac\Gamma\Theta\right),
\end{eqnarray}
where $\Theta\equiv3H$, $\gamma\equiv w+1$ ($w$ being the EoS parameter), and $\Gamma$ being the matter creation rate. Notice that the evolution of the deceleration parameter is influenced by the choice of the matter creation rate, $\Gamma$, and one can therefore explore various possibilities.  With the choice of 
the matter creation rate $\Gamma$ in the form of the 3-parameter decomposition (\ref{eq3_67}), the deceleration parameter takes on the form:
\begin{eqnarray}\label{deceleration}
q(H)=\frac{(\Gamma_1-3\delta)H^2+\Gamma_0H+\Gamma_{-1}}{3H^2(\delta-1)},
\end{eqnarray}
where $\delta=1-\frac{2}{3\gamma}$. 
Therefore, the transient acceleration condition $q(H)=0$ is equivalent to the existence of two distinct positive roots of the equation
\begin{eqnarray}\label{acc-transient}
(\Gamma_1-3\delta)H^2+\Gamma_0H+\Gamma_{-1}=0.
\end{eqnarray}
Now, for any arbitrary values of the model parameters, $\Gamma_i$'s, one can depict the evolution of the deceleration parameter from eqn. (\ref{deceleration}) and consequently check the existence of the transient acceleration from eqn. (\ref{acc-transient}). It is clear that the existence of two positive roots of eqn. (\ref{acc-transient}) depends on the values of $\Gamma_i$'s in the parametrized form of the matter creation rate. For instance, looking at eqn. (\ref{acc-transient}), one can derive that $\Gamma_1 \neq 3 \delta$ is a necessary condition for having two roots irrespective of its real or complex values and two distinct real roots are ensured for $\Gamma_0^2 - 4 (\Gamma_1-3\delta) \Gamma_{-1} > 0$. Thus, one can explore the nature of the equation (\ref{acc-transient}) in terms of the model parameters and consequently bound them in order to get the real roots. However, we comment that one could also start with a more complicated matter creation rate and investigate the equation $q(H) = 0$ in a similar fashion.  Now, as the model parameters, namely, $\Gamma_i$'s of the matter creation rate can be expressed in terms of the cosmographic parameters (see eqn. (\ref{eq3_69})), therefore, one can estimate the observational bounds on $\Gamma_i$'s through the estimations of $q_0, j_0, s_0$ (see section \ref{sec-constraints})  and try to depict the evolution of the deceleration parameter using this information. This is an indirect method to constrain the matter creation model (eqn. (\ref{eq3_69})) using the kinematical constraints.

In Fig. \ref{Fig-deceleration1},  we show the evolution of the deceleration parameter, $q(H)$,  using some fixed typical values of $q_0= -0.5, j_0= 1$ but with different values of the snap parameter $s_0$. The red, green and blue curve of  Fig. \ref{Fig-deceleration1} respectively stands for $s_0 = -10$, $-1.1$ and $-0.5$. In the same figure (i.e. Fig. \ref{Fig-deceleration1}), the  black dashed curve represents the evolution of the deceleration parameter for the SCM model, given by \[q_{\rm SCM}(H)=\frac12\left(1-3\frac{\Omega_0}{H^2}\right)\] with $\Omega_0\; (=0.73)$ being the density parameter of the total fluid at $z= 0$. From Fig. \ref{Fig-deceleration1}, one can see that for $s_0 = -10$ (red curve) and $s_0 = -1.1$ (green curve),
transient acceleration  is allowed while the blue curve drawn for $s_0 =-0.5$ only predicts the transition from the past decelerating phase to the present accelerating phase and this continues so. 
The SCM model also indicates an eternal acceleration after the decelerating phase (see the black dashed curve of Fig. \ref{Fig-deceleration1}). Therefore, one can see that depending on the values of the snap parameter, one could realize either an eternal accelerating phase or a transient feature in the deceleration parameter. In fact, from Fig. \ref{Fig-deceleration1}, one can further notice that the determination of the future deceleration of the universe is strongly dependent on the magnitude of the snap parameter provided $q_0$ and $j_0$ are fixed as done in this case.

In contrary to Fig. \ref{Fig-deceleration1} where we just fixed $q_0$ and $j_0$ but varied $s_0$ to understand the evolution of the deceleration parameter, in Fig. \ref{Fig-deceleration2}, we have depicted the evolution of the same parameter where we use the values of $(q_0, j_0, s_0)$ estimated from the observational datasets summarized in Table~\ref{tab:cosmography}. We see from Fig. \ref{Fig-deceleration2} that, for all the datasets shown in Table~\ref{tab:cosmography}, the parametrized form of the matter creation rate (\ref{eq3_67}) allows a decelerating phase of the universe in near future for some specific values of the model parameters. Since the matter creation theory has such interesting outcome, it will be equally interesting to extend this idea in other scenarios.

Usually two different approaches are used to depict the evolution of the deceleration parameter. In the first approach, one can introduce a cosmological model, such as the $\Lambda$CDM, $w$CDM, interacting dark energy, matter creation model, or any other cosmological model and then derive the expression of the deceleration parameter for each aforementioned model. On the other hand, one can directly parametrize the deceleration parameter $q $ without any cosmological model a priori, that means one can parametrize the deceleration parameter in terms of the scale factor or the redshift. This latter approach has also been investigated widely in the last couple of years, see Refs. \cite{Shapiro:2005nz,Gong:2006gs,Cunha:2008ja,Xu:2009zza,Santos:2010gp,2011PhLB..699..246L,delCampo:2012ya,Giostri:2012ek,Roman-Garza:2018cxf} for a variety of simple and complicated parametrizations of the deceleration parameter.  In this work, we consider three well known dark energy models, namely, $\Lambda$CDM; $w$CDM, where $w$ is the dark energy equation of state and it is a constant; $w(z)$CDM, where the dark energy equation of state $w (z)$ is variable having the form $w (z) = w_0 + w_a\; \frac{z}{1+z}$ (known as Chevallier-Polarski-Linder parametrization) \cite{Chevallier:2000qy,Linder:2002et} with $w_0 = w(z=0)$ and $w_a$ being two free parameters (we label this model as  $w_0w_a$CDM); along with 
following three typical parametrizations: $q_{I}(z) = q_0 + q_1\; z (1+z)^{-1}$ \cite{2011PhLB..699..246L}, $q_{II}(z) =   -1 + 1.5(1+z)^{q_2}/(q_1 + (1+z)^{q_2})$ \cite{delCampo:2012ya}, $q_{III}(z) = 0.5 + (q_0 - 0.5) (1+z) \exp\left(z_c^2 - (z+z_c)^2 \right)$ \cite{Roman-Garza:2018cxf}
where $q_0 = q(z=0)$, $q_1$, $q_2$ and $z_c$ are all free parameters.  
Additionally, along with the above approaches, we have also considered the expression of the deceleration parameter derived using  the cosmographic approach. In the cosmographic method,  
the expansion of the scale factor around the present time, $t_0$, can be expressed as  \cite{Visser2004}:

\begin{eqnarray}\label{series-scale-factor}
a (t) = 1 + H_0 (t-t_0) -\frac{1}{2} q_0 H_0^2 (t-t_0)^2
+ \frac{1}{3!} j_0 H_0^3 (t-t_0)^3  + \frac{1}{4!} s_0 H_0^4 (t-t_0)^4 + \mathcal{O} [(t-t_0)^5], 
\end{eqnarray}
where $H_0, q_0, j_0, s_0$ are the present values of the respective parameters defined in section \ref{sec-cosmographic-method}. Now using the series expansion of the scale factor (\ref{series-scale-factor}), one can write down the series expansion of the deceleration parameter in terms of the redshift $z$ as \cite{Guimaraes:2010mw}:
\begin{align}\label{q-model-independent}
q(z) = q_0 + (-q_0-2q_0^2+j_0)z + (2q_0+8q_0^2+8q_0^3-7q_0j_0-4j_0-s_0)\frac{z^2}{2}+ \mathcal{O} (z^3),
\end{align}
and this expression is not related to any underlying model. Thus, from its evolution one could determine the nature of the deceleration parameter from the model independent approach since the determination of the cosmographic parameters does not depend on any underlying cosmological model. Thus, overall, one can trace the evolution of the deceleration parameter using (i) any cosmological model such as $\Lambda$CDM, $w$CDM and some others as well, (ii) any parametrized form of the deceleration parameter, and finally (ii) from the cosmographic expression of the deceleration parameter given in eqn.  (\ref{q-model-independent}).

In Fig. \ref{Fig-deceleration3} we present the evolution of the deceleration parameter considering the above cosmological frameworks, that means, the known cosmological models (i.e. $\Lambda$CDM, $w$CDM, $w_0w_a$CDM), three different parametrizations of $q(z)$, namely, $q_{I}$, $q_{II}$, $q_{III}$ described above, and finally the cosmographic expression of the deceleration parameter given in eqn.  
(\ref{q-model-independent}). We note that while drawing the curves for $q(z)$ for $\Lambda$CDM, $w$CDM and $w_0w_a$CDM, the matter density parameter at present $\Omega_{m0}$ has been fixed to $0.3$ (consequently, the present value of the dark energy density parameter in these three models is, $1-\Omega_{m0} =0.7$). For $w$CDM and $w_0w_a$CDM models, we respectively fix
$w =-0.95$, and $(w_0, w_a) = (-0.65, -1.05)$ as obtained using the combined analysis Planck 2018+BAO in Ref. \cite{Yang:2021flj}. Concerning three different parametrizations of $q (z)$, we have taken the available values of the model parameters as in the literature\footnote{We note that for some parametrizations the constraints are presented using the old SNIa data. }. For $q_{I} (z)$, we take the mean values of $(q_0, q_1) = (-0.701, 1.718)$ obtained using Union2+$H(z)$  \cite{2011PhLB..699..246L}; for $q_{II} (z)$, we use the mean values of $(q_1, q_2) = (2.87, 3.27)$ using Union2+BAO/CMB+$H(z)$ from \cite{delCampo:2012ya}; for $q_{III} (z)$, we take the mean values of $(q_0, z_c) = (-0.51, 0.61)$ obtained using the analysis of Pantheon+$H(z)$ (from cosmic chronometers) as in Ref. \cite{Roman-Garza:2018cxf}. Finally, for the cosmographic expression of the deceleration parameter (eqn. (\ref{q-model-independent})), we take the mean values of ($q_0, j_0, s_0$) from Pantheon+CC (fifth column of Table~\ref{tab:cosmography}) and Pantheon+HST+CC (last column of Table~\ref{tab:cosmography}). 
From Fig. \ref{Fig-deceleration3}, one can see that the nature of the deceleration parameter is significantly dependent on the cosmological models and the parametrizations. For example, within $\Lambda$CDM (red dashed curve) and $w$CDM (green solid curve), we see that the universe enters the present accelerating phase from the past decelerating expansion and continues its accelerating expansion. While for the $w_0w_a$CDM model (black dotted curve), we see that in future there is a possibility of a decelerating expansion. On the other hand, looking at the parametrized forms of the deceleration parameter, we see that, for $q_{I}$ (brown dash double-dot curve) and $q_{II}$ (cyan double-dash dot curve), the eternal accelerating phase after the decelerating one continues.  The parametrization $q_{III}$ (orange dash-long dash curve),  shows a future deceleration. Finally,  the cosmographic expression of the deceleration parameter (eqn. (\ref{q-model-independent}))  also predicts a deceleration in the future.

Thus, one can clearly see that a possible decelerating expansion of the universe is highly dependent on the nature of the cosmological models and parametrizations. It is true that the matter creation model can allow such possibility for some specific choices of the model parameters but we cannot say that the evidence of a future deceleration is strong in this case. It is expected that the future cosmological surveys might offer more information about this phenomenon. 

\section{Conclusion}
\label{sec-conclu}

With the growing sensitivity of the observational data, cosmology has become more promising at present time. Looking at its theoretical developments together with the large amount of observational data, it has been argued that our universe should have experienced two accelerating phases, one during its very early evolution, known as inflation  and the latter is the late-time accelerating phase. Between these accelerating phases, a decelerating phase should have existed in agreement with the structure formation of the universe. This is a very rough picture of the universe based on the available information. Over the years, several attempts have been made to understand the complete dynamics of the universe, however, despite of many investigations, the complete dynamics of the universe is not yet known, and therefore, so many questions are left in the room.  
A very natural question that is closely related to the dynamics of the universe is the following: {\it whether our universe will continue with this accelerating expansion or it may decelerate in the near future.} This is a very justified question  because the possibility of a future deceleration of our universe may constrain abruptly the outlook of the modern cosmological picture and consequently a number of cosmological models might be ruled out.  According to a set of observational data, the indication for the slowing down of the cosmic acceleration was reported by several authors, mostly based on the dark energy models \cite{Starobinsky2009,Cardenas:2013roa,Magana:2014voa,Shahalam:2015lra,Hu:2015ksa}.   Being motivated with this question, in the present article, we have therefore made an attempt to investigate this issue in the context of the matter creation theory.  As already discussed, the theory of matter creation plays a crucial role in explaining various phases of the universe including the early accelerating phase, the intermediate matter dominated era and the present accelerating expansion of the universe, see the Refs.  \cite{Gunzig1998,Lima2007,Steigman:2008bc,Lima:2012cm,Pan2013,Lima:2014qpa,Ramos2014,Lima2016, Chakraborty2014,Chakraborty14,Pan2016,deHaro:2015hdp}.

Whilst investigating such delicate issue, we have introduced the cosmographic approach since one can find a connection between the cosmographic parameters and the free parameters of a cosmological model. Therefore, using the observational bounds on the cosmographic parameters, one can therefore find a rough estimations of the model parameters. 
Now considering a generalized matter creation rate we have presented the relation between the cosmographic parameters namely, $H_0$, $q_0$, $j_0$ and $s_0$ and the free parameters of the generalized matter creation rate. Then we have investigated the evolution of the deceleration parameter for the generalized matter creation theory using some arbitrary values of cosmographic parameters (i.e. for some arbitrary values of the model parameters since cosmographic parameters are related to the model parameters), see Fig. \ref{Fig-deceleration1} and then using the observational estimations of the cosmographic parameters from Table \ref{tab:cosmography}, see Fig. \ref{Fig-deceleration2}. From both Figs. \ref{Fig-deceleration1} and \ref{Fig-deceleration2}, 
we find that the generalized matter creation model may allow the possibility of a decelerating expansion of the universe in future, but, this possibility depends on many factors, for instance the observational data used to constrain the cosmographic parameters play a very important role in this direction.

However, the theory of matter creation is not the only one predicting the possibility of the transient acceleration in late universe. In fact, according to the existing literature, the possibility of such transient acceleration is not a new result in cosmology. Earlier investigations in several cosmological theories have pointed out that the recent accelerating phase of our universe may not be eternal and the expansion of our universe may decelerate again in future, see for  instance a list of works performed by several investigators in various directions with  similar conclusion \cite{Russo:2004ym,Bilic:2005sp,Carvalho:2006fy,Srivastava:2006xq,Wu:2008sc,Bento:2008yx,Fabris:2009mn,Guimaraes:2010mw,Bose:2010gc,Vargas:2011sz,Pan2013,Chakraborty2014,Chen:2011cy,Pan:2014lua,Qi:2014yxa}.  Nevertheless, this possibility depends highly on the background cosmological model or parametrization. In Fig. \ref{Fig-deceleration3}, we have investigated this issue considering  a number of well known cosmological models, some phenomenological parametrizations of the deceleration parameter, and a cosmographic expression of the deceleration parameter where we see that both eternal acceleration  as well as a future deceleration of our universe are allowed.

Based on the present examinations, it is clear that even though the possibility of the future deceleration is allowed in the context of matter creation model (also in some other models and parametrizations) for some specific values of the model parameters, but we cannot claim that the evidence for a future deceleration of our universe is strong, at least at this stage. 
We certainly believe that the investigations toward this direction should be performed with more potential cosmological datasets from the upcoming cosmological surveys and it is expected that future cosmological surveys will  shed more light in this direction.
Lastly, we conclude that the aim of the present article is not to focus only on the possible decelerating expansion of the universe, rather, to be more informative about the dynamics of the universe in the future. 

\section{Acknowledgments}
The authors are thankful to the referees for some important comments that helped us to improve the quality of the manuscript.
SP was supported by the Mathematical Research Impact-Centric Support Scheme (MATRICS), File No. MTR/2018/000940, given by the Science and Engineering Research Board (SERB), Govt. of India. WY was supported by the  National Natural Science Foundation of China under Grants No. 11705079 and No. 11647153.

\bibliography{bibliofinal.bib}

\providecommand{\noopsort}[1]{}\providecommand{\singleletter}[1]{#1}%
\begin{thebibliography}{128}
\expandafter\ifx\csname natexlab\endcsname\relax\def\natexlab#1{#1}\fi
\expandafter\ifx\csname bibnamefont\endcsname\relax
  \def\bibnamefont#1{#1}\fi
\expandafter\ifx\csname bibfnamefont\endcsname\relax
  \def\bibfnamefont#1{#1}\fi
\expandafter\ifx\csname citenamefont\endcsname\relax
  \def\citenamefont#1{#1}\fi
\expandafter\ifx\csname url\endcsname\relax
  \def\url#1{\texttt{#1}}\fi
\expandafter\ifx\csname urlprefix\endcsname\relax\def\urlprefix{URL }\fi
\providecommand{\bibinfo}[2]{#2}
\providecommand{\eprint}[2][]{\url{#2}}

\bibitem[{\citenamefont{Riess et~al.}(1998)}]{Riess:1998cb}
\bibinfo{author}{\bibfnamefont{A.~G.} \bibnamefont{Riess}} \bibnamefont{et~al.}
  (\bibinfo{collaboration}{Supernova Search Team}), \bibinfo{journal}{Astron.
  J.} \textbf{\bibinfo{volume}{116}}, \bibinfo{pages}{1009}
  (\bibinfo{year}{1998}), \eprint{astro-ph/9805201}.

\bibitem[{\citenamefont{Perlmutter et~al.}(1999)}]{Perlmutter:1998np}
\bibinfo{author}{\bibfnamefont{S.}~\bibnamefont{Perlmutter}}
  \bibnamefont{et~al.} (\bibinfo{collaboration}{Supernova Cosmology Project}),
  \bibinfo{journal}{Astrophys. J.} \textbf{\bibinfo{volume}{517}},
  \bibinfo{pages}{565} (\bibinfo{year}{1999}), \eprint{astro-ph/9812133}.

\bibitem[{\citenamefont{Hinshaw et~al.}(2013)}]{Hinshaw:2012aka}
\bibinfo{author}{\bibfnamefont{G.}~\bibnamefont{Hinshaw}} \bibnamefont{et~al.}
  (\bibinfo{collaboration}{WMAP}), \bibinfo{journal}{Astrophys. J. Suppl.}
  \textbf{\bibinfo{volume}{208}}, \bibinfo{pages}{19} (\bibinfo{year}{2013}),
  \eprint{1212.5226}.

\bibitem[{\citenamefont{Suzuki et~al.}(2012)}]{Suzuki:2011hu}
\bibinfo{author}{\bibfnamefont{N.}~\bibnamefont{Suzuki}} \bibnamefont{et~al.}
  (\bibinfo{collaboration}{Supernova Cosmology Project}),
  \bibinfo{journal}{Astrophys. J.} \textbf{\bibinfo{volume}{746}},
  \bibinfo{pages}{85} (\bibinfo{year}{2012}), \eprint{1105.3470}.

\bibitem[{\citenamefont{Aghanim et~al.}(2020)}]{Planck:2018vyg}
\bibinfo{author}{\bibfnamefont{N.}~\bibnamefont{Aghanim}} \bibnamefont{et~al.}
  (\bibinfo{collaboration}{Planck}), \bibinfo{journal}{Astron. Astrophys.}
  \textbf{\bibinfo{volume}{641}}, \bibinfo{pages}{A6} (\bibinfo{year}{2020}),
  \bibinfo{note}{[Erratum: Astron.Astrophys. 652, C4 (2021)]},
  \eprint{1807.06209}.

\bibitem[{\citenamefont{Starobinsky}(1979)}]{Starobinsky:1979ty}
\bibinfo{author}{\bibfnamefont{A.~A.} \bibnamefont{Starobinsky}},
  \bibinfo{journal}{JETP Lett.} \textbf{\bibinfo{volume}{30}},
  \bibinfo{pages}{682} (\bibinfo{year}{1979}).

\bibitem[{\citenamefont{Starobinsky}(1980)}]{Starobinsky:1980te}
\bibinfo{author}{\bibfnamefont{A.~A.} \bibnamefont{Starobinsky}},
  \bibinfo{journal}{Phys. Lett. B} \textbf{\bibinfo{volume}{91}},
  \bibinfo{pages}{99} (\bibinfo{year}{1980}).

\bibitem[{\citenamefont{Guth}(1981)}]{Guth:1980zm}
\bibinfo{author}{\bibfnamefont{A.~H.} \bibnamefont{Guth}},
  \bibinfo{journal}{Phys. Rev. D} \textbf{\bibinfo{volume}{23}},
  \bibinfo{pages}{347} (\bibinfo{year}{1981}).

\bibitem[{\citenamefont{Linde}(1982)}]{Linde:1981mu}
\bibinfo{author}{\bibfnamefont{A.~D.} \bibnamefont{Linde}},
  \bibinfo{journal}{Adv. Ser. Astrophys. Cosmol.}
  \textbf{\bibinfo{volume}{108}}, \bibinfo{pages}{389} (\bibinfo{year}{1982}).

\bibitem[{\citenamefont{Mukhanov and Chibisov}(1981)}]{Mukhanov:1981xt}
\bibinfo{author}{\bibfnamefont{V.~F.} \bibnamefont{Mukhanov}} \bibnamefont{and}
  \bibinfo{author}{\bibfnamefont{G.~V.} \bibnamefont{Chibisov}},
  \bibinfo{journal}{JETP Lett.} \textbf{\bibinfo{volume}{33}},
  \bibinfo{pages}{532} (\bibinfo{year}{1981}).

\bibitem[{\citenamefont{Barrow and Turner}(1981)}]{Barrow:1981pa}
\bibinfo{author}{\bibfnamefont{J.~D.} \bibnamefont{Barrow}} \bibnamefont{and}
  \bibinfo{author}{\bibfnamefont{M.~S.} \bibnamefont{Turner}},
  \bibinfo{journal}{Nature} \textbf{\bibinfo{volume}{292}}, \bibinfo{pages}{35}
  (\bibinfo{year}{1981}).

\bibitem[{\citenamefont{Hawking}(1982)}]{Hawking:1982cz}
\bibinfo{author}{\bibfnamefont{S.~W.} \bibnamefont{Hawking}},
  \bibinfo{journal}{Phys. Lett. B} \textbf{\bibinfo{volume}{115}},
  \bibinfo{pages}{295} (\bibinfo{year}{1982}).

\bibitem[{\citenamefont{Starobinsky}(1982)}]{Starobinsky:1982ee}
\bibinfo{author}{\bibfnamefont{A.~A.} \bibnamefont{Starobinsky}},
  \bibinfo{journal}{Phys. Lett. B} \textbf{\bibinfo{volume}{117}},
  \bibinfo{pages}{175} (\bibinfo{year}{1982}).

\bibitem[{\citenamefont{Guth and Pi}(1982)}]{Guth:1982ec}
\bibinfo{author}{\bibfnamefont{A.~H.} \bibnamefont{Guth}} \bibnamefont{and}
  \bibinfo{author}{\bibfnamefont{S.}~\bibnamefont{Pi}}, \bibinfo{journal}{Phys.
  Rev. Lett.} \textbf{\bibinfo{volume}{49}}, \bibinfo{pages}{1110}
  (\bibinfo{year}{1982}).

\bibitem[{\citenamefont{Starobinsky}(1983)}]{Starobinsky:1983zz}
\bibinfo{author}{\bibfnamefont{A.~A.} \bibnamefont{Starobinsky}},
  \bibinfo{journal}{Sov. Astron. Lett.} \textbf{\bibinfo{volume}{9}},
  \bibinfo{pages}{302} (\bibinfo{year}{1983}).

\bibitem[{\citenamefont{Barrow and Turner}(1982)}]{Barrow:1982kr}
\bibinfo{author}{\bibfnamefont{J.~D.} \bibnamefont{Barrow}} \bibnamefont{and}
  \bibinfo{author}{\bibfnamefont{M.~S.} \bibnamefont{Turner}},
  \bibinfo{journal}{Nature} \textbf{\bibinfo{volume}{298}},
  \bibinfo{pages}{801} (\bibinfo{year}{1982}).

\bibitem[{\citenamefont{Linde}(1984)}]{Linde:1984ir}
\bibinfo{author}{\bibfnamefont{A.~D.} \bibnamefont{Linde}},
  \bibinfo{journal}{Rept. Prog. Phys.} \textbf{\bibinfo{volume}{47}},
  \bibinfo{pages}{925} (\bibinfo{year}{1984}).

\bibitem[{\citenamefont{Linde}(1985)}]{Linde:1985ub}
\bibinfo{author}{\bibfnamefont{A.~D.} \bibnamefont{Linde}},
  \bibinfo{journal}{Phys. Lett. B} \textbf{\bibinfo{volume}{162}},
  \bibinfo{pages}{281} (\bibinfo{year}{1985}).

\bibitem[{\citenamefont{Kofman and Linde}(1987)}]{Kofman:1986wm}
\bibinfo{author}{\bibfnamefont{L.}~\bibnamefont{Kofman}} \bibnamefont{and}
  \bibinfo{author}{\bibfnamefont{A.~D.} \bibnamefont{Linde}},
  \bibinfo{journal}{Nucl. Phys. B} \textbf{\bibinfo{volume}{282}},
  \bibinfo{pages}{555} (\bibinfo{year}{1987}).

\bibitem[{\citenamefont{Burd and Barrow}(1988)}]{Burd:1988ss}
\bibinfo{author}{\bibfnamefont{A.}~\bibnamefont{Burd}} \bibnamefont{and}
  \bibinfo{author}{\bibfnamefont{J.~D.} \bibnamefont{Barrow}},
  \bibinfo{journal}{Nucl. Phys. B} \textbf{\bibinfo{volume}{308}},
  \bibinfo{pages}{929} (\bibinfo{year}{1988}).

\bibitem[{\citenamefont{Barrow}(1990)}]{Barrow:1990vx}
\bibinfo{author}{\bibfnamefont{J.~D.} \bibnamefont{Barrow}},
  \bibinfo{journal}{Phys. Lett. B} \textbf{\bibinfo{volume}{235}},
  \bibinfo{pages}{40} (\bibinfo{year}{1990}).

\bibitem[{\citenamefont{Barrow}(1993)}]{Barrow:1993hn}
\bibinfo{author}{\bibfnamefont{J.~D.} \bibnamefont{Barrow}},
  \bibinfo{journal}{Phys. Rev. D} \textbf{\bibinfo{volume}{48}},
  \bibinfo{pages}{1585} (\bibinfo{year}{1993}).

\bibitem[{\citenamefont{Barrow and Parsons}(1995)}]{Barrow:1995xb}
\bibinfo{author}{\bibfnamefont{J.~D.} \bibnamefont{Barrow}} \bibnamefont{and}
  \bibinfo{author}{\bibfnamefont{P.}~\bibnamefont{Parsons}},
  \bibinfo{journal}{Phys. Rev. D} \textbf{\bibinfo{volume}{52}},
  \bibinfo{pages}{5576} (\bibinfo{year}{1995}), \eprint{astro-ph/9506049}.

\bibitem[{\citenamefont{Barrow and Paliathanasis}(2016)}]{Barrow:2016qkh}
\bibinfo{author}{\bibfnamefont{J.~D.} \bibnamefont{Barrow}} \bibnamefont{and}
  \bibinfo{author}{\bibfnamefont{A.}~\bibnamefont{Paliathanasis}},
  \bibinfo{journal}{Phys. Rev. D} \textbf{\bibinfo{volume}{94}},
  \bibinfo{pages}{083518} (\bibinfo{year}{2016}), \eprint{1609.01126}.

\bibitem[{\citenamefont{Paliathanasis}(2017{\natexlab{a}})}]{Paliathanasis:2017apr}
\bibinfo{author}{\bibfnamefont{A.}~\bibnamefont{Paliathanasis}},
  \bibinfo{journal}{Eur. Phys. J. C} \textbf{\bibinfo{volume}{77}},
  \bibinfo{pages}{438} (\bibinfo{year}{2017}{\natexlab{a}}),
  \eprint{1706.06400}.

\bibitem[{\citenamefont{Peebles and Vilenkin}(1999)}]{Peebles:1998qn}
\bibinfo{author}{\bibfnamefont{P.}~\bibnamefont{Peebles}} \bibnamefont{and}
  \bibinfo{author}{\bibfnamefont{A.}~\bibnamefont{Vilenkin}},
  \bibinfo{journal}{Phys. Rev. D} \textbf{\bibinfo{volume}{59}},
  \bibinfo{pages}{063505} (\bibinfo{year}{1999}), \eprint{astro-ph/9810509}.

\bibitem[{\citenamefont{Banerjee and Pav\'on}(2001)}]{Banerjee:2000mj}
\bibinfo{author}{\bibfnamefont{N.}~\bibnamefont{Banerjee}} \bibnamefont{and}
  \bibinfo{author}{\bibfnamefont{D.}~\bibnamefont{Pav\'on}},
  \bibinfo{journal}{Phys. Rev. D} \textbf{\bibinfo{volume}{63}},
  \bibinfo{pages}{043504} (\bibinfo{year}{2001}), \eprint{gr-qc/0012048}.

\bibitem[{\citenamefont{Peebles and Ratra}(2003)}]{Peebles:2002gy}
\bibinfo{author}{\bibfnamefont{P.}~\bibnamefont{Peebles}} \bibnamefont{and}
  \bibinfo{author}{\bibfnamefont{B.}~\bibnamefont{Ratra}},
  \bibinfo{journal}{Rev. Mod. Phys.} \textbf{\bibinfo{volume}{75}},
  \bibinfo{pages}{559} (\bibinfo{year}{2003}), \eprint{astro-ph/0207347}.

\bibitem[{\citenamefont{Pan and Chakraborty}(2013{\natexlab{a}})}]{Pan:2013rha}
\bibinfo{author}{\bibfnamefont{S.}~\bibnamefont{Pan}} \bibnamefont{and}
  \bibinfo{author}{\bibfnamefont{S.}~\bibnamefont{Chakraborty}},
  \bibinfo{journal}{Eur. Phys. J. C} \textbf{\bibinfo{volume}{73}},
  \bibinfo{pages}{2575} (\bibinfo{year}{2013}{\natexlab{a}}),
  \eprint{1303.5602}.

\bibitem[{\citenamefont{Paliathanasis
  et~al.}(2015{\natexlab{a}})\citenamefont{Paliathanasis, Tsamparlis,
  Basilakos, and Barrow}}]{Paliathanasis:2015gga}
\bibinfo{author}{\bibfnamefont{A.}~\bibnamefont{Paliathanasis}},
  \bibinfo{author}{\bibfnamefont{M.}~\bibnamefont{Tsamparlis}},
  \bibinfo{author}{\bibfnamefont{S.}~\bibnamefont{Basilakos}},
  \bibnamefont{and} \bibinfo{author}{\bibfnamefont{J.~D.}
  \bibnamefont{Barrow}}, \bibinfo{journal}{Phys. Rev. D}
  \textbf{\bibinfo{volume}{91}}, \bibinfo{pages}{123535}
  (\bibinfo{year}{2015}{\natexlab{a}}), \eprint{1503.05750}.

\bibitem[{\citenamefont{Paliathanasis
  et~al.}(2015{\natexlab{b}})\citenamefont{Paliathanasis, Pan, and
  Pramanik}}]{Paliathanasis:2015cza}
\bibinfo{author}{\bibfnamefont{A.}~\bibnamefont{Paliathanasis}},
  \bibinfo{author}{\bibfnamefont{S.}~\bibnamefont{Pan}}, \bibnamefont{and}
  \bibinfo{author}{\bibfnamefont{S.}~\bibnamefont{Pramanik}},
  \bibinfo{journal}{Class. Quant. Grav.} \textbf{\bibinfo{volume}{32}},
  \bibinfo{pages}{245006} (\bibinfo{year}{2015}{\natexlab{b}}),
  \eprint{1508.06543}.

\bibitem[{\citenamefont{de~Haro
  et~al.}(2016{\natexlab{a}})\citenamefont{de~Haro, Amor\'os, and
  Pan}}]{deHaro:2016hpl}
\bibinfo{author}{\bibfnamefont{J.}~\bibnamefont{de~Haro}},
  \bibinfo{author}{\bibfnamefont{J.}~\bibnamefont{Amor\'os}}, \bibnamefont{and}
  \bibinfo{author}{\bibfnamefont{S.}~\bibnamefont{Pan}},
  \bibinfo{journal}{Phys. Rev. D} \textbf{\bibinfo{volume}{93}},
  \bibinfo{pages}{084018} (\bibinfo{year}{2016}{\natexlab{a}}),
  \eprint{1601.08175}.

\bibitem[{\citenamefont{de~Haro
  et~al.}(2016{\natexlab{b}})\citenamefont{de~Haro, Amor\'os, and
  Pan}}]{deHaro:2016cdm}
\bibinfo{author}{\bibfnamefont{J.}~\bibnamefont{de~Haro}},
  \bibinfo{author}{\bibfnamefont{J.}~\bibnamefont{Amor\'os}}, \bibnamefont{and}
  \bibinfo{author}{\bibfnamefont{S.}~\bibnamefont{Pan}},
  \bibinfo{journal}{Phys. Rev. D} \textbf{\bibinfo{volume}{94}},
  \bibinfo{pages}{064060} (\bibinfo{year}{2016}{\natexlab{b}}),
  \eprint{1607.06726}.

\bibitem[{\citenamefont{Paliathanasis}(2017{\natexlab{b}})}]{Paliathanasis:2017efk}
\bibinfo{author}{\bibfnamefont{A.}~\bibnamefont{Paliathanasis}},
  \bibinfo{journal}{Phys. Rev. D} \textbf{\bibinfo{volume}{95}},
  \bibinfo{pages}{064062} (\bibinfo{year}{2017}{\natexlab{b}}),
  \eprint{1701.04360}.

\bibitem[{\citenamefont{Yang et~al.}(2017)\citenamefont{Yang, Banerjee, and
  Pan}}]{Yang:2017yme}
\bibinfo{author}{\bibfnamefont{W.}~\bibnamefont{Yang}},
  \bibinfo{author}{\bibfnamefont{N.}~\bibnamefont{Banerjee}}, \bibnamefont{and}
  \bibinfo{author}{\bibfnamefont{S.}~\bibnamefont{Pan}},
  \bibinfo{journal}{Phys. Rev. D} \textbf{\bibinfo{volume}{95}},
  \bibinfo{pages}{123527} (\bibinfo{year}{2017}), \eprint{1705.09278}.

\bibitem[{\citenamefont{Das et~al.}(2018{\natexlab{a}})\citenamefont{Das,
  Banerjee, Chakraborty, and Pan}}]{Das:2017gjj}
\bibinfo{author}{\bibfnamefont{A.}~\bibnamefont{Das}},
  \bibinfo{author}{\bibfnamefont{A.}~\bibnamefont{Banerjee}},
  \bibinfo{author}{\bibfnamefont{S.}~\bibnamefont{Chakraborty}},
  \bibnamefont{and} \bibinfo{author}{\bibfnamefont{S.}~\bibnamefont{Pan}},
  \bibinfo{journal}{Pramana} \textbf{\bibinfo{volume}{90}}, \bibinfo{pages}{19}
  (\bibinfo{year}{2018}{\natexlab{a}}), \eprint{1706.08145}.

\bibitem[{\citenamefont{Yang et~al.}(2018{\natexlab{a}})\citenamefont{Yang,
  Pan, Di~Valentino, Nunes, Vagnozzi, and Mota}}]{Yang:2018euj}
\bibinfo{author}{\bibfnamefont{W.}~\bibnamefont{Yang}},
  \bibinfo{author}{\bibfnamefont{S.}~\bibnamefont{Pan}},
  \bibinfo{author}{\bibfnamefont{E.}~\bibnamefont{Di~Valentino}},
  \bibinfo{author}{\bibfnamefont{R.~C.} \bibnamefont{Nunes}},
  \bibinfo{author}{\bibfnamefont{S.}~\bibnamefont{Vagnozzi}}, \bibnamefont{and}
  \bibinfo{author}{\bibfnamefont{D.~F.} \bibnamefont{Mota}},
  \bibinfo{journal}{JCAP} \textbf{\bibinfo{volume}{09}}, \bibinfo{pages}{019}
  (\bibinfo{year}{2018}{\natexlab{a}}), \eprint{1805.08252}.

\bibitem[{\citenamefont{Pan et~al.}(2018)\citenamefont{Pan, Mukherjee, and
  Banerjee}}]{Pan:2017ent}
\bibinfo{author}{\bibfnamefont{S.}~\bibnamefont{Pan}},
  \bibinfo{author}{\bibfnamefont{A.}~\bibnamefont{Mukherjee}},
  \bibnamefont{and} \bibinfo{author}{\bibfnamefont{N.}~\bibnamefont{Banerjee}},
  \bibinfo{journal}{Mon. Not. Roy. Astron. Soc.}
  \textbf{\bibinfo{volume}{477}}, \bibinfo{pages}{1189} (\bibinfo{year}{2018}),
  \eprint{1710.03725}.

\bibitem[{\citenamefont{Das et~al.}(2018{\natexlab{b}})\citenamefont{Das, Pan,
  Ghosh, and Pal}}]{Das:2018bzx}
\bibinfo{author}{\bibfnamefont{P.}~\bibnamefont{Das}},
  \bibinfo{author}{\bibfnamefont{S.}~\bibnamefont{Pan}},
  \bibinfo{author}{\bibfnamefont{S.}~\bibnamefont{Ghosh}}, \bibnamefont{and}
  \bibinfo{author}{\bibfnamefont{P.}~\bibnamefont{Pal}},
  \bibinfo{journal}{Phys. Rev. D} \textbf{\bibinfo{volume}{98}},
  \bibinfo{pages}{024004} (\bibinfo{year}{2018}{\natexlab{b}}),
  \eprint{1801.07970}.

\bibitem[{\citenamefont{Barrow and Paliathanasis}(2018)}]{Barrow:2016wiy}
\bibinfo{author}{\bibfnamefont{J.~D.} \bibnamefont{Barrow}} \bibnamefont{and}
  \bibinfo{author}{\bibfnamefont{A.}~\bibnamefont{Paliathanasis}},
  \bibinfo{journal}{Gen. Rel. Grav.} \textbf{\bibinfo{volume}{50}},
  \bibinfo{pages}{82} (\bibinfo{year}{2018}), \eprint{1611.06680}.

\bibitem[{\citenamefont{Yang et~al.}(2018{\natexlab{b}})\citenamefont{Yang,
  Pan, and Barrow}}]{Yang:2017zjs}
\bibinfo{author}{\bibfnamefont{W.}~\bibnamefont{Yang}},
  \bibinfo{author}{\bibfnamefont{S.}~\bibnamefont{Pan}}, \bibnamefont{and}
  \bibinfo{author}{\bibfnamefont{J.~D.} \bibnamefont{Barrow}},
  \bibinfo{journal}{Phys. Rev. D} \textbf{\bibinfo{volume}{97}},
  \bibinfo{pages}{043529} (\bibinfo{year}{2018}{\natexlab{b}}),
  \eprint{1706.04953}.

\bibitem[{\citenamefont{Haro et~al.}(2019)\citenamefont{Haro, Amor\'os, and
  Pan}}]{Haro:2019gsv}
\bibinfo{author}{\bibfnamefont{J.}~\bibnamefont{Haro}},
  \bibinfo{author}{\bibfnamefont{J.}~\bibnamefont{Amor\'os}}, \bibnamefont{and}
  \bibinfo{author}{\bibfnamefont{S.}~\bibnamefont{Pan}}, \bibinfo{journal}{Eur.
  Phys. J. C} \textbf{\bibinfo{volume}{79}}, \bibinfo{pages}{505}
  (\bibinfo{year}{2019}), \eprint{1901.00167}.

\bibitem[{\citenamefont{Yang et~al.}(2019{\natexlab{a}})\citenamefont{Yang,
  Banerjee, Paliathanasis, and Pan}}]{Yang:2018qec}
\bibinfo{author}{\bibfnamefont{W.}~\bibnamefont{Yang}},
  \bibinfo{author}{\bibfnamefont{N.}~\bibnamefont{Banerjee}},
  \bibinfo{author}{\bibfnamefont{A.}~\bibnamefont{Paliathanasis}},
  \bibnamefont{and} \bibinfo{author}{\bibfnamefont{S.}~\bibnamefont{Pan}},
  \bibinfo{journal}{Phys. Dark Univ.} \textbf{\bibinfo{volume}{26}},
  \bibinfo{pages}{100383} (\bibinfo{year}{2019}{\natexlab{a}}),
  \eprint{1812.06854}.

\bibitem[{\citenamefont{Yang et~al.}(2019{\natexlab{b}})\citenamefont{Yang,
  Pan, Paliathanasis, Ghosh, and Wu}}]{Yang:2019jwn}
\bibinfo{author}{\bibfnamefont{W.}~\bibnamefont{Yang}},
  \bibinfo{author}{\bibfnamefont{S.}~\bibnamefont{Pan}},
  \bibinfo{author}{\bibfnamefont{A.}~\bibnamefont{Paliathanasis}},
  \bibinfo{author}{\bibfnamefont{S.}~\bibnamefont{Ghosh}}, \bibnamefont{and}
  \bibinfo{author}{\bibfnamefont{Y.}~\bibnamefont{Wu}}, \bibinfo{journal}{Mon.
  Not. Roy. Astron. Soc.} \textbf{\bibinfo{volume}{490}}, \bibinfo{pages}{2071}
  (\bibinfo{year}{2019}{\natexlab{b}}), \eprint{1904.10436}.

\bibitem[{\citenamefont{Yang et~al.}(2019{\natexlab{c}})\citenamefont{Yang,
  Pan, Di~Valentino, Paliathanasis, and Lu}}]{Yang:2019qza}
\bibinfo{author}{\bibfnamefont{W.}~\bibnamefont{Yang}},
  \bibinfo{author}{\bibfnamefont{S.}~\bibnamefont{Pan}},
  \bibinfo{author}{\bibfnamefont{E.}~\bibnamefont{Di~Valentino}},
  \bibinfo{author}{\bibfnamefont{A.}~\bibnamefont{Paliathanasis}},
  \bibnamefont{and} \bibinfo{author}{\bibfnamefont{J.}~\bibnamefont{Lu}},
  \bibinfo{journal}{Phys. Rev. D} \textbf{\bibinfo{volume}{100}},
  \bibinfo{pages}{103518} (\bibinfo{year}{2019}{\natexlab{c}}),
  \eprint{1906.04162}.

\bibitem[{\citenamefont{Paliathanasis
  et~al.}(2019{\natexlab{a}})\citenamefont{Paliathanasis, Pan, and
  Yang}}]{Paliathanasis:2019hbi}
\bibinfo{author}{\bibfnamefont{A.}~\bibnamefont{Paliathanasis}},
  \bibinfo{author}{\bibfnamefont{S.}~\bibnamefont{Pan}}, \bibnamefont{and}
  \bibinfo{author}{\bibfnamefont{W.}~\bibnamefont{Yang}},
  \bibinfo{journal}{Int. J. Mod. Phys. D} \textbf{\bibinfo{volume}{28}},
  \bibinfo{pages}{1950161} (\bibinfo{year}{2019}{\natexlab{a}}),
  \eprint{1903.02370}.

\bibitem[{\citenamefont{Das et~al.}(2019)\citenamefont{Das, Pan, and
  Ghosh}}]{Das:2018ylw}
\bibinfo{author}{\bibfnamefont{P.}~\bibnamefont{Das}},
  \bibinfo{author}{\bibfnamefont{S.}~\bibnamefont{Pan}}, \bibnamefont{and}
  \bibinfo{author}{\bibfnamefont{S.}~\bibnamefont{Ghosh}},
  \bibinfo{journal}{Phys. Lett. B} \textbf{\bibinfo{volume}{791}},
  \bibinfo{pages}{66} (\bibinfo{year}{2019}), \eprint{1810.06606}.

\bibitem[{\citenamefont{Paliathanasis
  et~al.}(2019{\natexlab{b}})\citenamefont{Paliathanasis, Leon, and
  Pan}}]{Paliathanasis:2018vru}
\bibinfo{author}{\bibfnamefont{A.}~\bibnamefont{Paliathanasis}},
  \bibinfo{author}{\bibfnamefont{G.}~\bibnamefont{Leon}}, \bibnamefont{and}
  \bibinfo{author}{\bibfnamefont{S.}~\bibnamefont{Pan}}, \bibinfo{journal}{Gen.
  Rel. Grav.} \textbf{\bibinfo{volume}{51}}, \bibinfo{pages}{106}
  (\bibinfo{year}{2019}{\natexlab{b}}), \eprint{1811.10038}.

\bibitem[{\citenamefont{Haro et~al.}(2020)\citenamefont{Haro, Amor\'os, and
  Pan}}]{Haro:2019peq}
\bibinfo{author}{\bibfnamefont{J.}~\bibnamefont{Haro}},
  \bibinfo{author}{\bibfnamefont{J.}~\bibnamefont{Amor\'os}}, \bibnamefont{and}
  \bibinfo{author}{\bibfnamefont{S.}~\bibnamefont{Pan}}, \bibinfo{journal}{Eur.
  Phys. J. C} \textbf{\bibinfo{volume}{80}}, \bibinfo{pages}{404}
  (\bibinfo{year}{2020}), \eprint{1908.01516}.

\bibitem[{\citenamefont{Giacomini et~al.}(2020)\citenamefont{Giacomini, Leon,
  Paliathanasis, and Pan}}]{Giacomini:2020grc}
\bibinfo{author}{\bibfnamefont{A.}~\bibnamefont{Giacomini}},
  \bibinfo{author}{\bibfnamefont{G.}~\bibnamefont{Leon}},
  \bibinfo{author}{\bibfnamefont{A.}~\bibnamefont{Paliathanasis}},
  \bibnamefont{and} \bibinfo{author}{\bibfnamefont{S.}~\bibnamefont{Pan}},
  \bibinfo{journal}{Eur. Phys. J. C} \textbf{\bibinfo{volume}{80}},
  \bibinfo{pages}{184} (\bibinfo{year}{2020}), \eprint{2001.02414}.

\bibitem[{\citenamefont{Pan et~al.}(2020)\citenamefont{Pan, de~Haro, Yang, and
  Amor\'os}}]{Pan:2020mst}
\bibinfo{author}{\bibfnamefont{S.}~\bibnamefont{Pan}},
  \bibinfo{author}{\bibfnamefont{J.}~\bibnamefont{de~Haro}},
  \bibinfo{author}{\bibfnamefont{W.}~\bibnamefont{Yang}}, \bibnamefont{and}
  \bibinfo{author}{\bibfnamefont{J.}~\bibnamefont{Amor\'os}},
  \bibinfo{journal}{2001.09885}  (\bibinfo{year}{2020}).

\bibitem[{\citenamefont{Sahni and Starobinsky}(2000)}]{Sahni:1999gb}
\bibinfo{author}{\bibfnamefont{V.}~\bibnamefont{Sahni}} \bibnamefont{and}
  \bibinfo{author}{\bibfnamefont{A.~A.} \bibnamefont{Starobinsky}},
  \bibinfo{journal}{Int. J. Mod. Phys. D} \textbf{\bibinfo{volume}{9}},
  \bibinfo{pages}{373} (\bibinfo{year}{2000}), \eprint{astro-ph/9904398}.

\bibitem[{\citenamefont{Sahni and Starobinsky}(2006)}]{Sahni:2006pa}
\bibinfo{author}{\bibfnamefont{V.}~\bibnamefont{Sahni}} \bibnamefont{and}
  \bibinfo{author}{\bibfnamefont{A.}~\bibnamefont{Starobinsky}},
  \bibinfo{journal}{Int. J. Mod. Phys. D} \textbf{\bibinfo{volume}{15}},
  \bibinfo{pages}{2105} (\bibinfo{year}{2006}), \eprint{astro-ph/0610026}.

\bibitem[{\citenamefont{Copeland et~al.}(2006)\citenamefont{Copeland, Sami, and
  Tsujikawa}}]{Copeland:2006wr}
\bibinfo{author}{\bibfnamefont{E.~J.} \bibnamefont{Copeland}},
  \bibinfo{author}{\bibfnamefont{M.}~\bibnamefont{Sami}}, \bibnamefont{and}
  \bibinfo{author}{\bibfnamefont{S.}~\bibnamefont{Tsujikawa}},
  \bibinfo{journal}{Int. J. Mod. Phys. D} \textbf{\bibinfo{volume}{15}},
  \bibinfo{pages}{1753} (\bibinfo{year}{2006}), \eprint{hep-th/0603057}.

\bibitem[{\citenamefont{Di~Valentino et~al.}(2021)\citenamefont{Di~Valentino,
  Mena, Pan, Visinelli, Yang, Melchiorri, Mota, Riess, and
  Silk}}]{DiValentino:2021izs}
\bibinfo{author}{\bibfnamefont{E.}~\bibnamefont{Di~Valentino}},
  \bibinfo{author}{\bibfnamefont{O.}~\bibnamefont{Mena}},
  \bibinfo{author}{\bibfnamefont{S.}~\bibnamefont{Pan}},
  \bibinfo{author}{\bibfnamefont{L.}~\bibnamefont{Visinelli}},
  \bibinfo{author}{\bibfnamefont{W.}~\bibnamefont{Yang}},
  \bibinfo{author}{\bibfnamefont{A.}~\bibnamefont{Melchiorri}},
  \bibinfo{author}{\bibfnamefont{D.~F.} \bibnamefont{Mota}},
  \bibinfo{author}{\bibfnamefont{A.~G.} \bibnamefont{Riess}}, \bibnamefont{and}
  \bibinfo{author}{\bibfnamefont{J.}~\bibnamefont{Silk}},
  \bibinfo{journal}{Class. Quant. Grav.} \textbf{\bibinfo{volume}{38}},
  \bibinfo{pages}{153001} (\bibinfo{year}{2021}), \eprint{2103.01183}.

\bibitem[{\citenamefont{Perivolaropoulos and
  Skara}(2021)}]{Perivolaropoulos:2021jda}
\bibinfo{author}{\bibfnamefont{L.}~\bibnamefont{Perivolaropoulos}}
  \bibnamefont{and} \bibinfo{author}{\bibfnamefont{F.}~\bibnamefont{Skara}}
  (\bibinfo{year}{2021}), \eprint{2105.05208}.

\bibitem[{\citenamefont{Steigman et~al.}(2009)\citenamefont{Steigman, Santos,
  and Lima}}]{Steigman:2008bc}
\bibinfo{author}{\bibfnamefont{G.}~\bibnamefont{Steigman}},
  \bibinfo{author}{\bibfnamefont{R.}~\bibnamefont{Santos}}, \bibnamefont{and}
  \bibinfo{author}{\bibfnamefont{J.}~\bibnamefont{Lima}},
  \bibinfo{journal}{JCAP} \textbf{\bibinfo{volume}{06}}, \bibinfo{pages}{033}
  (\bibinfo{year}{2009}), \eprint{0812.3912}.

\bibitem[{\citenamefont{Lima et~al.}(2012)\citenamefont{Lima, Basilakos, and
  Costa}}]{Lima:2012cm}
\bibinfo{author}{\bibfnamefont{J.}~\bibnamefont{Lima}},
  \bibinfo{author}{\bibfnamefont{S.}~\bibnamefont{Basilakos}},
  \bibnamefont{and} \bibinfo{author}{\bibfnamefont{F.}~\bibnamefont{Costa}},
  \bibinfo{journal}{Phys. Rev. D} \textbf{\bibinfo{volume}{86}},
  \bibinfo{pages}{103534} (\bibinfo{year}{2012}), \eprint{1205.0868}.

\bibitem[{\citenamefont{Lima et~al.}(2014)\citenamefont{Lima, Graef, Pavon, and
  Basilakos}}]{Lima:2014qpa}
\bibinfo{author}{\bibfnamefont{J.}~\bibnamefont{Lima}},
  \bibinfo{author}{\bibfnamefont{L.}~\bibnamefont{Graef}},
  \bibinfo{author}{\bibfnamefont{D.}~\bibnamefont{Pavon}}, \bibnamefont{and}
  \bibinfo{author}{\bibfnamefont{S.}~\bibnamefont{Basilakos}},
  \bibinfo{journal}{JCAP} \textbf{\bibinfo{volume}{10}}, \bibinfo{pages}{042}
  (\bibinfo{year}{2014}), \eprint{1406.5538}.

\bibitem[{\citenamefont{Ramos et~al.}(2014)\citenamefont{Ramos, Vargas~dos
  Santos, and Waga}}]{Ramos2014}
\bibinfo{author}{\bibfnamefont{R.~O.} \bibnamefont{Ramos}},
  \bibinfo{author}{\bibfnamefont{M.}~\bibnamefont{Vargas~dos Santos}},
  \bibnamefont{and} \bibinfo{author}{\bibfnamefont{I.}~\bibnamefont{Waga}},
  \bibinfo{journal}{Phys. Rev. D} \textbf{\bibinfo{volume}{89}},
  \bibinfo{pages}{083524} (\bibinfo{year}{2014}), \eprint{arXiv:1404.2604}.

\bibitem[{\citenamefont{Lima et~al.}(2016)\citenamefont{Lima, Santos, and
  Cunha}}]{Lima2016}
\bibinfo{author}{\bibfnamefont{J.~A.~S.} \bibnamefont{Lima}},
  \bibinfo{author}{\bibfnamefont{R.~C.} \bibnamefont{Santos}},
  \bibnamefont{and} \bibinfo{author}{\bibfnamefont{J.~V.} \bibnamefont{Cunha}},
  \bibinfo{journal}{JCAP} \textbf{\bibinfo{volume}{2016}}, \bibinfo{pages}{027}
  (\bibinfo{year}{2016}), \eprint{arXiv:1508.07263}.

\bibitem[{\citenamefont{Pan et~al.}(2016)\citenamefont{Pan, de~Haro,
  Paliathanasis, and Slagter}}]{Pan2016}
\bibinfo{author}{\bibfnamefont{S.}~\bibnamefont{Pan}},
  \bibinfo{author}{\bibfnamefont{J.}~\bibnamefont{de~Haro}},
  \bibinfo{author}{\bibfnamefont{A.}~\bibnamefont{Paliathanasis}},
  \bibnamefont{and} \bibinfo{author}{\bibfnamefont{R.~J.}
  \bibnamefont{Slagter}}, \bibinfo{journal}{Mon. Not. Roy. Astron. Soc.}
  \textbf{\bibinfo{volume}{460}}, \bibinfo{pages}{1445} (\bibinfo{year}{2016}),
  \eprint{arXiv:1601.03955}.

\bibitem[{\citenamefont{Gunzig et~al.}(1998)\citenamefont{Gunzig, Maartens, and
  Nesteruk}}]{Gunzig1998}
\bibinfo{author}{\bibfnamefont{E.}~\bibnamefont{Gunzig}},
  \bibinfo{author}{\bibfnamefont{R.}~\bibnamefont{Maartens}}, \bibnamefont{and}
  \bibinfo{author}{\bibfnamefont{A.~V.} \bibnamefont{Nesteruk}},
  \bibinfo{journal}{Classical and Quantum Gravity}
  \textbf{\bibinfo{volume}{15}}, \bibinfo{pages}{923} (\bibinfo{year}{1998}),
  \eprint{arXiv:astro-ph/9703137}.

\bibitem[{\citenamefont{Abramo and Lima}(1996)}]{Abramo:1996ip}
\bibinfo{author}{\bibfnamefont{L.}~\bibnamefont{Abramo}} \bibnamefont{and}
  \bibinfo{author}{\bibfnamefont{J.}~\bibnamefont{Lima}},
  \bibinfo{journal}{Class. Quant. Grav.} \textbf{\bibinfo{volume}{13}},
  \bibinfo{pages}{2953} (\bibinfo{year}{1996}), \eprint{gr-qc/9606064}.

\bibitem[{\citenamefont{de~Haro and Pan}(2016)}]{deHaro:2015hdp}
\bibinfo{author}{\bibfnamefont{J.}~\bibnamefont{de~Haro}} \bibnamefont{and}
  \bibinfo{author}{\bibfnamefont{S.}~\bibnamefont{Pan}},
  \bibinfo{journal}{Class. Quant. Grav.} \textbf{\bibinfo{volume}{33}},
  \bibinfo{pages}{165007} (\bibinfo{year}{2016}), \eprint{1512.03100}.

\bibitem[{\citenamefont{Lima et~al.}(2013)\citenamefont{Lima, Basilakos, and
  Sola}}]{Lima:2012mu}
\bibinfo{author}{\bibfnamefont{J.}~\bibnamefont{Lima}},
  \bibinfo{author}{\bibfnamefont{S.}~\bibnamefont{Basilakos}},
  \bibnamefont{and} \bibinfo{author}{\bibfnamefont{J.}~\bibnamefont{Sola}},
  \bibinfo{journal}{Mon. Not. Roy. Astron. Soc.}
  \textbf{\bibinfo{volume}{431}}, \bibinfo{pages}{923} (\bibinfo{year}{2013}),
  \eprint{1209.2802}.

\bibitem[{\citenamefont{Perico et~al.}(2013)\citenamefont{Perico, Lima,
  Basilakos, and Sola}}]{Perico:2013mna}
\bibinfo{author}{\bibfnamefont{E.}~\bibnamefont{Perico}},
  \bibinfo{author}{\bibfnamefont{J.}~\bibnamefont{Lima}},
  \bibinfo{author}{\bibfnamefont{S.}~\bibnamefont{Basilakos}},
  \bibnamefont{and} \bibinfo{author}{\bibfnamefont{J.}~\bibnamefont{Sola}},
  \bibinfo{journal}{Phys. Rev. D} \textbf{\bibinfo{volume}{88}},
  \bibinfo{pages}{063531} (\bibinfo{year}{2013}), \eprint{1306.0591}.

\bibitem[{\citenamefont{Weinberg}(1972)}]{Weinberg1972}
\bibinfo{author}{\bibfnamefont{S.}~\bibnamefont{Weinberg}},
  \emph{\bibinfo{title}{Gravitation and Cosmology}} (\bibinfo{publisher}{John
  Wiley and Sons}, \bibinfo{address}{New York}, \bibinfo{year}{1972}).

\bibitem[{\citenamefont{Visser}(2005)}]{Visser2005}
\bibinfo{author}{\bibfnamefont{M.}~\bibnamefont{Visser}},
  \bibinfo{journal}{General Relativity and Gravitation}
  \textbf{\bibinfo{volume}{37}}, \bibinfo{pages}{1541} (\bibinfo{year}{2005}),
  \eprint{arXiv:gr-qc/0411131}.

\bibitem[{\citenamefont{Visser}(2004)}]{Visser2004}
\bibinfo{author}{\bibfnamefont{M.}~\bibnamefont{Visser}},
  \bibinfo{journal}{Classical and Quantum Gravity}
  \textbf{\bibinfo{volume}{21}}, \bibinfo{pages}{2603} (\bibinfo{year}{2004}),
  \eprint{arXiv:gr-qc/0309109}.

\bibitem[{\citenamefont{Dunajski and Gibbons}(2008)}]{Dunajski2008}
\bibinfo{author}{\bibfnamefont{M.}~\bibnamefont{Dunajski}} \bibnamefont{and}
  \bibinfo{author}{\bibfnamefont{G.}~\bibnamefont{Gibbons}},
  \bibinfo{journal}{Classical and Quantum Gravity}
  \textbf{\bibinfo{volume}{25}}, \bibinfo{pages}{235012}
  (\bibinfo{year}{2008}), \eprint{arXiv:0807.0207}.

\bibitem[{\citenamefont{Guimar\~{a}es and Lima}(2011)}]{Guimaraes:2010mw}
\bibinfo{author}{\bibfnamefont{A.~C.} \bibnamefont{Guimar\~{a}es}}
  \bibnamefont{and} \bibinfo{author}{\bibfnamefont{J.~A.~S.}
  \bibnamefont{Lima}}, \bibinfo{journal}{Class. Quant. Grav.}
  \textbf{\bibinfo{volume}{28}}, \bibinfo{pages}{125026}
  (\bibinfo{year}{2011}), \eprint{1005.2986}.

\bibitem[{\citenamefont{Bolotin et~al.}(2012)\citenamefont{Bolotin, Lemets, and
  Yerokhin}}]{Bolotin2012}
\bibinfo{author}{\bibfnamefont{Y.~L.} \bibnamefont{Bolotin}},
  \bibinfo{author}{\bibfnamefont{O.~A.} \bibnamefont{Lemets}},
  \bibnamefont{and} \bibinfo{author}{\bibfnamefont{D.~A.}
  \bibnamefont{Yerokhin}}, \bibinfo{journal}{Physics-Uspekhi}
  \textbf{\bibinfo{volume}{55}}, \bibinfo{pages}{876} (\bibinfo{year}{2012}),
  \eprint{arXiv:1511.06532}.

\bibitem[{\citenamefont{Bolotin et~al.}(2016)\citenamefont{Bolotin, Cherkaskiy,
  and Lemets}}]{Bolotin2016}
\bibinfo{author}{\bibfnamefont{Y.~L.} \bibnamefont{Bolotin}},
  \bibinfo{author}{\bibfnamefont{V.~A.} \bibnamefont{Cherkaskiy}},
  \bibnamefont{and} \bibinfo{author}{\bibfnamefont{O.~A.}
  \bibnamefont{Lemets}}, \bibinfo{journal}{International Journal of Modern
  Physics D} \textbf{\bibinfo{volume}{25}}, \bibinfo{pages}{1650056}
  (\bibinfo{year}{2016}), \eprint{arXiv:1503.04056}.

\bibitem[{\citenamefont{Dunsby and Luongo}(2016)}]{Dunsby2016}
\bibinfo{author}{\bibfnamefont{P.~K.~S.} \bibnamefont{Dunsby}}
  \bibnamefont{and} \bibinfo{author}{\bibfnamefont{O.}~\bibnamefont{Luongo}},
  \bibinfo{journal}{International Journal of Geometric Methods in Modern
  Physics} \textbf{\bibinfo{volume}{13}}, \bibinfo{pages}{1630002}
  (\bibinfo{year}{2016}), \eprint{arXiv:1511.06532}.

\bibitem[{\citenamefont{Bolotin et~al.}(2018)\citenamefont{Bolotin, Cherkaskiy,
  Ivashtenko, Konchatnyi, and Zazunov}}]{Bolotin2018}
\bibinfo{author}{\bibfnamefont{Y.~L.} \bibnamefont{Bolotin}},
  \bibinfo{author}{\bibfnamefont{V.~A.} \bibnamefont{Cherkaskiy}},
  \bibinfo{author}{\bibfnamefont{O.~Y.} \bibnamefont{Ivashtenko}},
  \bibinfo{author}{\bibfnamefont{M.~I.} \bibnamefont{Konchatnyi}},
  \bibnamefont{and} \bibinfo{author}{\bibfnamefont{L.~G.}
  \bibnamefont{Zazunov}} (\bibinfo{year}{2018}), \eprint{1812.02394}.

\bibitem[{\citenamefont{Sandage}(1962)}]{Sandage1962}
\bibinfo{author}{\bibfnamefont{A.}~\bibnamefont{Sandage}},
  \bibinfo{journal}{Astrophysical Journal} \textbf{\bibinfo{volume}{136}},
  \bibinfo{pages}{319} (\bibinfo{year}{1962}).

\bibitem[{\citenamefont{Sahni et~al.}(2003)\citenamefont{Sahni, Saini,
  Starobinsky, and Alam}}]{Sahni:2002fz}
\bibinfo{author}{\bibfnamefont{V.}~\bibnamefont{Sahni}},
  \bibinfo{author}{\bibfnamefont{T.~D.} \bibnamefont{Saini}},
  \bibinfo{author}{\bibfnamefont{A.~A.} \bibnamefont{Starobinsky}},
  \bibnamefont{and} \bibinfo{author}{\bibfnamefont{U.}~\bibnamefont{Alam}},
  \bibinfo{journal}{JETP Lett.} \textbf{\bibinfo{volume}{77}},
  \bibinfo{pages}{201} (\bibinfo{year}{2003}), \eprint{astro-ph/0201498}.

\bibitem[{\citenamefont{Alam et~al.}(2003)\citenamefont{Alam, Sahni, Saini, and
  Starobinsky}}]{Alam:2003sc}
\bibinfo{author}{\bibfnamefont{U.}~\bibnamefont{Alam}},
  \bibinfo{author}{\bibfnamefont{V.}~\bibnamefont{Sahni}},
  \bibinfo{author}{\bibfnamefont{T.~D.} \bibnamefont{Saini}}, \bibnamefont{and}
  \bibinfo{author}{\bibfnamefont{A.~A.} \bibnamefont{Starobinsky}},
  \bibinfo{journal}{Mon. Not. Roy. Astron. Soc.}
  \textbf{\bibinfo{volume}{344}}, \bibinfo{pages}{1057} (\bibinfo{year}{2003}),
  \eprint{astro-ph/0303009}.

\bibitem[{\citenamefont{Arabsalmani and Sahni}(2011)}]{Arabsalmani:2011fz}
\bibinfo{author}{\bibfnamefont{M.}~\bibnamefont{Arabsalmani}} \bibnamefont{and}
  \bibinfo{author}{\bibfnamefont{V.}~\bibnamefont{Sahni}},
  \bibinfo{journal}{Phys. Rev. D} \textbf{\bibinfo{volume}{83}},
  \bibinfo{pages}{043501} (\bibinfo{year}{2011}), \eprint{1101.3436}.

\bibitem[{\citenamefont{Moresco et~al.}(2016)\citenamefont{Moresco, Pozzetti,
  Cimatti, Jimenez, Maraston, Verde, Thomas, Citro, Tojeiro, and
  Wilkinson}}]{Moresco:2016mzx}
\bibinfo{author}{\bibfnamefont{M.}~\bibnamefont{Moresco}},
  \bibinfo{author}{\bibfnamefont{L.}~\bibnamefont{Pozzetti}},
  \bibinfo{author}{\bibfnamefont{A.}~\bibnamefont{Cimatti}},
  \bibinfo{author}{\bibfnamefont{R.}~\bibnamefont{Jimenez}},
  \bibinfo{author}{\bibfnamefont{C.}~\bibnamefont{Maraston}},
  \bibinfo{author}{\bibfnamefont{L.}~\bibnamefont{Verde}},
  \bibinfo{author}{\bibfnamefont{D.}~\bibnamefont{Thomas}},
  \bibinfo{author}{\bibfnamefont{A.}~\bibnamefont{Citro}},
  \bibinfo{author}{\bibfnamefont{R.}~\bibnamefont{Tojeiro}}, \bibnamefont{and}
  \bibinfo{author}{\bibfnamefont{D.}~\bibnamefont{Wilkinson}},
  \bibinfo{journal}{JCAP} \textbf{\bibinfo{volume}{2016}}, \bibinfo{pages}{014}
  (\bibinfo{year}{2016}).

\bibitem[{\citenamefont{Scolnic et~al.}(2018)}]{Scolnic:2017caz}
\bibinfo{author}{\bibfnamefont{D.}~\bibnamefont{Scolnic}} \bibnamefont{et~al.},
  \bibinfo{journal}{Astrophys. J.} \textbf{\bibinfo{volume}{859}},
  \bibinfo{pages}{101} (\bibinfo{year}{2018}), \eprint{1710.00845}.

\bibitem[{\citenamefont{Riess et~al.}(2019)\citenamefont{Riess, Casertano,
  Yuan, Macri, and Scolnic}}]{Riess:2019cxk}
\bibinfo{author}{\bibfnamefont{A.~G.} \bibnamefont{Riess}},
  \bibinfo{author}{\bibfnamefont{S.}~\bibnamefont{Casertano}},
  \bibinfo{author}{\bibfnamefont{W.}~\bibnamefont{Yuan}},
  \bibinfo{author}{\bibfnamefont{L.~M.} \bibnamefont{Macri}}, \bibnamefont{and}
  \bibinfo{author}{\bibfnamefont{D.}~\bibnamefont{Scolnic}},
  \bibinfo{journal}{Astrophys. J.} \textbf{\bibinfo{volume}{876}},
  \bibinfo{pages}{85} (\bibinfo{year}{2019}), \eprint{1903.07603}.

\bibitem[{\citenamefont{Cattoen and Visser}(2007)}]{Cattoen:2007sk}
\bibinfo{author}{\bibfnamefont{C.}~\bibnamefont{Cattoen}} \bibnamefont{and}
  \bibinfo{author}{\bibfnamefont{M.}~\bibnamefont{Visser}},
  \bibinfo{journal}{Class. Quant. Grav.} \textbf{\bibinfo{volume}{24}},
  \bibinfo{pages}{5985} (\bibinfo{year}{2007}), \eprint{0710.1887}.

\bibitem[{\citenamefont{Lewis and Bridle}(2002)}]{Lewis:2002ah}
\bibinfo{author}{\bibfnamefont{A.}~\bibnamefont{Lewis}} \bibnamefont{and}
  \bibinfo{author}{\bibfnamefont{S.}~\bibnamefont{Bridle}},
  \bibinfo{journal}{Phys. Rev. D} \textbf{\bibinfo{volume}{66}},
  \bibinfo{pages}{103511} (\bibinfo{year}{2002}), \eprint{astro-ph/0205436}.

\bibitem[{\citenamefont{Lewis et~al.}(2000)\citenamefont{Lewis, Challinor, and
  Lasenby}}]{Lewis:1999bs}
\bibinfo{author}{\bibfnamefont{A.}~\bibnamefont{Lewis}},
  \bibinfo{author}{\bibfnamefont{A.}~\bibnamefont{Challinor}},
  \bibnamefont{and} \bibinfo{author}{\bibfnamefont{A.}~\bibnamefont{Lasenby}},
  \bibinfo{journal}{Astrophys. J.} \textbf{\bibinfo{volume}{538}},
  \bibinfo{pages}{473} (\bibinfo{year}{2000}), \eprint{astro-ph/9911177}.

\bibitem[{\citenamefont{Gelman and Rubin}(1992)}]{Gelman-Rubin}
\bibinfo{author}{\bibfnamefont{A.}~\bibnamefont{Gelman}} \bibnamefont{and}
  \bibinfo{author}{\bibfnamefont{D.}~\bibnamefont{Rubin}},
  \bibinfo{journal}{Statistical Science} \textbf{\bibinfo{volume}{7}},
  \bibinfo{pages}{457} (\bibinfo{year}{1992}).

\bibitem[{\citenamefont{{Xu} and {Wang}}(2011)}]{2011PhLB..702..114X}
\bibinfo{author}{\bibfnamefont{L.}~\bibnamefont{{Xu}}} \bibnamefont{and}
  \bibinfo{author}{\bibfnamefont{Y.}~\bibnamefont{{Wang}}},
  \bibinfo{journal}{Physics Letters B} \textbf{\bibinfo{volume}{702}},
  \bibinfo{pages}{114} (\bibinfo{year}{2011}), \eprint{1009.0963}.

\bibitem[{\citenamefont{Li et~al.}(2020)\citenamefont{Li, Du, and
  Xu}}]{Li:2019qic}
\bibinfo{author}{\bibfnamefont{E.-K.} \bibnamefont{Li}},
  \bibinfo{author}{\bibfnamefont{M.}~\bibnamefont{Du}}, \bibnamefont{and}
  \bibinfo{author}{\bibfnamefont{L.}~\bibnamefont{Xu}}, \bibinfo{journal}{Mon.
  Not. Roy. Astron. Soc.} \textbf{\bibinfo{volume}{491}}, \bibinfo{pages}{4960}
  (\bibinfo{year}{2020}), \eprint{1903.11433}.

\bibitem[{\citenamefont{Zel'dovich}(1970)}]{Zeldovich1970}
\bibinfo{author}{\bibfnamefont{Y.~B.} \bibnamefont{Zel'dovich}},
  \bibinfo{journal}{Soviet Journal of Experimental and Theoretical Physics
  Letters} \textbf{\bibinfo{volume}{12}}, \bibinfo{pages}{307}
  (\bibinfo{year}{1970}).

\bibitem[{\citenamefont{Prigogine}(1989)}]{Prigogine:1989wc}
\bibinfo{author}{\bibfnamefont{I.}~\bibnamefont{Prigogine}},
  \bibinfo{journal}{Int. J. Theor. Phys.} \textbf{\bibinfo{volume}{28}},
  \bibinfo{pages}{927} (\bibinfo{year}{1989}).

\bibitem[{\citenamefont{Lima et~al.}(2008)\citenamefont{Lima, Calvao, and
  Waga}}]{Lima2007}
\bibinfo{author}{\bibfnamefont{J.~A.~S.} \bibnamefont{Lima}},
  \bibinfo{author}{\bibfnamefont{M.~O.} \bibnamefont{Calvao}},
  \bibnamefont{and} \bibinfo{author}{\bibfnamefont{I.}~\bibnamefont{Waga}},
  \emph{\bibinfo{title}{Cosmology, thermodynamics and matter creation}}
  (\bibinfo{year}{2008}).

\bibitem[{\citenamefont{Pan and Chakraborty}(2013{\natexlab{b}})}]{Pan2013}
\bibinfo{author}{\bibfnamefont{S.}~\bibnamefont{Pan}} \bibnamefont{and}
  \bibinfo{author}{\bibfnamefont{S.}~\bibnamefont{Chakraborty}},
  \bibinfo{journal}{The European Physical Journal C}
  \textbf{\bibinfo{volume}{73}}, \bibinfo{pages}{2575}
  (\bibinfo{year}{2013}{\natexlab{b}}), \eprint{arXiv:1303.5602}.

\bibitem[{\citenamefont{Chakraborty et~al.}(2014)\citenamefont{Chakraborty,
  Pan, and Saha}}]{Chakraborty2014}
\bibinfo{author}{\bibfnamefont{S.}~\bibnamefont{Chakraborty}},
  \bibinfo{author}{\bibfnamefont{S.}~\bibnamefont{Pan}}, \bibnamefont{and}
  \bibinfo{author}{\bibfnamefont{S.}~\bibnamefont{Saha}},
  \bibinfo{journal}{Phys. Lett. B} \textbf{\bibinfo{volume}{738}},
  \bibinfo{pages}{424} (\bibinfo{year}{2014}), \eprint{arXiv:1411.0941}.

\bibitem[{\citenamefont{Chakraborty}(2014)}]{Chakraborty14}
\bibinfo{author}{\bibfnamefont{S.}~\bibnamefont{Chakraborty}},
  \bibinfo{journal}{Phys. Lett. B} \textbf{\bibinfo{volume}{732}},
  \bibinfo{pages}{81} (\bibinfo{year}{2014}), \eprint{arXiv:1403.5980}.

\bibitem[{\citenamefont{Pan and Chakraborty}(2015)}]{Pan:2014lua}
\bibinfo{author}{\bibfnamefont{S.}~\bibnamefont{Pan}} \bibnamefont{and}
  \bibinfo{author}{\bibfnamefont{S.}~\bibnamefont{Chakraborty}},
  \bibinfo{journal}{Adv. High Energy Phys.} \textbf{\bibinfo{volume}{2015}},
  \bibinfo{pages}{654025} (\bibinfo{year}{2015}), \eprint{1404.3273}.

\bibitem[{\citenamefont{Zeldovich and Starobinsky}(1971)}]{Zeldovich:1971mw}
\bibinfo{author}{\bibfnamefont{Y.~B.} \bibnamefont{Zeldovich}}
  \bibnamefont{and} \bibinfo{author}{\bibfnamefont{A.~A.}
  \bibnamefont{Starobinsky}}, \bibinfo{journal}{Zh. Eksp. Teor. Fiz.}
  \textbf{\bibinfo{volume}{61}}, \bibinfo{pages}{2161} (\bibinfo{year}{1971}).

\bibitem[{\citenamefont{Shafieloo et~al.}(2018)\citenamefont{Shafieloo, Hazra,
  Sahni, and Starobinsky}}]{Shafieloo:2016bpk}
\bibinfo{author}{\bibfnamefont{A.}~\bibnamefont{Shafieloo}},
  \bibinfo{author}{\bibfnamefont{D.~K.} \bibnamefont{Hazra}},
  \bibinfo{author}{\bibfnamefont{V.}~\bibnamefont{Sahni}}, \bibnamefont{and}
  \bibinfo{author}{\bibfnamefont{A.~A.} \bibnamefont{Starobinsky}},
  \bibinfo{journal}{Mon. Not. Roy. Astron. Soc.}
  \textbf{\bibinfo{volume}{473}}, \bibinfo{pages}{2760} (\bibinfo{year}{2018}),
  \eprint{1610.05192}.

\bibitem[{\citenamefont{Shafieloo et~al.}(2009)\citenamefont{Shafieloo, Sahni,
  and Starobinsky}}]{Starobinsky2009}
\bibinfo{author}{\bibfnamefont{A.}~\bibnamefont{Shafieloo}},
  \bibinfo{author}{\bibfnamefont{V.}~\bibnamefont{Sahni}}, \bibnamefont{and}
  \bibinfo{author}{\bibfnamefont{A.~A.} \bibnamefont{Starobinsky}},
  \bibinfo{journal}{Phys. Rev. D} \textbf{\bibinfo{volume}{80}},
  \bibinfo{pages}{101301} (\bibinfo{year}{2009}).

\bibitem[{\citenamefont{Cardenas et~al.}(2013)\citenamefont{Cardenas, Bernal,
  and Bonilla}}]{Cardenas:2013roa}
\bibinfo{author}{\bibfnamefont{V.~H.} \bibnamefont{Cardenas}},
  \bibinfo{author}{\bibfnamefont{C.}~\bibnamefont{Bernal}}, \bibnamefont{and}
  \bibinfo{author}{\bibfnamefont{A.}~\bibnamefont{Bonilla}},
  \bibinfo{journal}{Mon. Not. Roy. Astron. Soc.}
  \textbf{\bibinfo{volume}{433}}, \bibinfo{pages}{3534} (\bibinfo{year}{2013}),
  \eprint{1306.0779}.

\bibitem[{\citenamefont{Maga\~na et~al.}(2014)\citenamefont{Maga\~na,
  C\'ardenas, and Motta}}]{Magana:2014voa}
\bibinfo{author}{\bibfnamefont{J.}~\bibnamefont{Maga\~na}},
  \bibinfo{author}{\bibfnamefont{V.~H.} \bibnamefont{C\'ardenas}},
  \bibnamefont{and} \bibinfo{author}{\bibfnamefont{V.}~\bibnamefont{Motta}},
  \bibinfo{journal}{JCAP} \textbf{\bibinfo{volume}{10}}, \bibinfo{pages}{017}
  (\bibinfo{year}{2014}), \eprint{1407.1632}.

\bibitem[{\citenamefont{Shahalam et~al.}(2015)\citenamefont{Shahalam, Sami, and
  Agarwal}}]{Shahalam:2015lra}
\bibinfo{author}{\bibfnamefont{M.}~\bibnamefont{Shahalam}},
  \bibinfo{author}{\bibfnamefont{S.}~\bibnamefont{Sami}}, \bibnamefont{and}
  \bibinfo{author}{\bibfnamefont{A.}~\bibnamefont{Agarwal}},
  \bibinfo{journal}{Mon. Not. Roy. Astron. Soc.}
  \textbf{\bibinfo{volume}{448}}, \bibinfo{pages}{2948} (\bibinfo{year}{2015}),
  \eprint{1501.04047}.

\bibitem[{\citenamefont{Hu et~al.}(2016)\citenamefont{Hu, Li, Li, and
  Wang}}]{Hu:2015ksa}
\bibinfo{author}{\bibfnamefont{Y.}~\bibnamefont{Hu}},
  \bibinfo{author}{\bibfnamefont{M.}~\bibnamefont{Li}},
  \bibinfo{author}{\bibfnamefont{N.}~\bibnamefont{Li}}, \bibnamefont{and}
  \bibinfo{author}{\bibfnamefont{S.}~\bibnamefont{Wang}},
  \bibinfo{journal}{Astrophys. J.} \textbf{\bibinfo{volume}{821}},
  \bibinfo{pages}{60} (\bibinfo{year}{2016}), \eprint{1509.03461}.

\bibitem[{\citenamefont{Zhang and Xia}(2016)}]{Zhang:2016tto}
\bibinfo{author}{\bibfnamefont{M.-J.} \bibnamefont{Zhang}} \bibnamefont{and}
  \bibinfo{author}{\bibfnamefont{J.-Q.} \bibnamefont{Xia}},
  \bibinfo{journal}{JCAP} \textbf{\bibinfo{volume}{12}}, \bibinfo{pages}{005}
  (\bibinfo{year}{2016}), \eprint{1606.04398}.

\bibitem[{\citenamefont{Zhang and Xia}(2018)}]{Zhang:2017jvo}
\bibinfo{author}{\bibfnamefont{M.-J.} \bibnamefont{Zhang}} \bibnamefont{and}
  \bibinfo{author}{\bibfnamefont{J.-Q.} \bibnamefont{Xia}},
  \bibinfo{journal}{Nucl. Phys. B} \textbf{\bibinfo{volume}{929}},
  \bibinfo{pages}{438} (\bibinfo{year}{2018}), \eprint{1701.04973}.

\bibitem[{\citenamefont{Shapiro and Turner}(2006)}]{Shapiro:2005nz}
\bibinfo{author}{\bibfnamefont{C.}~\bibnamefont{Shapiro}} \bibnamefont{and}
  \bibinfo{author}{\bibfnamefont{M.~S.} \bibnamefont{Turner}},
  \bibinfo{journal}{Astrophys. J.} \textbf{\bibinfo{volume}{649}},
  \bibinfo{pages}{563} (\bibinfo{year}{2006}), \eprint{astro-ph/0512586}.

\bibitem[{\citenamefont{Gong and Wang}(2007)}]{Gong:2006gs}
\bibinfo{author}{\bibfnamefont{Y.-G.} \bibnamefont{Gong}} \bibnamefont{and}
  \bibinfo{author}{\bibfnamefont{A.}~\bibnamefont{Wang}},
  \bibinfo{journal}{Phys. Rev. D} \textbf{\bibinfo{volume}{75}},
  \bibinfo{pages}{043520} (\bibinfo{year}{2007}), \eprint{astro-ph/0612196}.

\bibitem[{\citenamefont{Cunha and Lima}(2008)}]{Cunha:2008ja}
\bibinfo{author}{\bibfnamefont{J.~V.} \bibnamefont{Cunha}} \bibnamefont{and}
  \bibinfo{author}{\bibfnamefont{J.~A.~S.} \bibnamefont{Lima}},
  \bibinfo{journal}{Mon. Not. Roy. Astron. Soc.}
  \textbf{\bibinfo{volume}{390}}, \bibinfo{pages}{210} (\bibinfo{year}{2008}),
  \eprint{0805.1261}.

\bibitem[{\citenamefont{Xu and Lu}(2009)}]{Xu:2009zza}
\bibinfo{author}{\bibfnamefont{L.}~\bibnamefont{Xu}} \bibnamefont{and}
  \bibinfo{author}{\bibfnamefont{J.}~\bibnamefont{Lu}}, \bibinfo{journal}{Mod.
  Phys. Lett. A} \textbf{\bibinfo{volume}{24}}, \bibinfo{pages}{369}
  (\bibinfo{year}{2009}).

\bibitem[{\citenamefont{Santos et~al.}(2011)\citenamefont{Santos, Carvalho, and
  Alcaniz}}]{Santos:2010gp}
\bibinfo{author}{\bibfnamefont{B.}~\bibnamefont{Santos}},
  \bibinfo{author}{\bibfnamefont{J.~C.} \bibnamefont{Carvalho}},
  \bibnamefont{and} \bibinfo{author}{\bibfnamefont{J.~S.}
  \bibnamefont{Alcaniz}}, \bibinfo{journal}{Astropart. Phys.}
  \textbf{\bibinfo{volume}{35}}, \bibinfo{pages}{17} (\bibinfo{year}{2011}),
  \eprint{1009.2733}.

\bibitem[{\citenamefont{{Lu} et~al.}(2011)\citenamefont{{Lu}, {Xu}, and
  {Liu}}}]{2011PhLB..699..246L}
\bibinfo{author}{\bibfnamefont{J.}~\bibnamefont{{Lu}}},
  \bibinfo{author}{\bibfnamefont{L.}~\bibnamefont{{Xu}}}, \bibnamefont{and}
  \bibinfo{author}{\bibfnamefont{M.}~\bibnamefont{{Liu}}},
  \bibinfo{journal}{Physics Letters B} \textbf{\bibinfo{volume}{699}},
  \bibinfo{pages}{246} (\bibinfo{year}{2011}), \eprint{1105.1871}.

\bibitem[{\citenamefont{del Campo et~al.}(2012)\citenamefont{del Campo, Duran,
  Herrera, and Pavon}}]{delCampo:2012ya}
\bibinfo{author}{\bibfnamefont{S.}~\bibnamefont{del Campo}},
  \bibinfo{author}{\bibfnamefont{I.}~\bibnamefont{Duran}},
  \bibinfo{author}{\bibfnamefont{R.}~\bibnamefont{Herrera}}, \bibnamefont{and}
  \bibinfo{author}{\bibfnamefont{D.}~\bibnamefont{Pavon}},
  \bibinfo{journal}{Phys. Rev. D} \textbf{\bibinfo{volume}{86}},
  \bibinfo{pages}{083509} (\bibinfo{year}{2012}), \eprint{1209.3415}.

\bibitem[{\citenamefont{Giostri et~al.}(2012)\citenamefont{Giostri, dos Santos,
  Waga, Reis, Calv\~ao, and Lago}}]{Giostri:2012ek}
\bibinfo{author}{\bibfnamefont{R.}~\bibnamefont{Giostri}},
  \bibinfo{author}{\bibfnamefont{M.~V.} \bibnamefont{dos Santos}},
  \bibinfo{author}{\bibfnamefont{I.}~\bibnamefont{Waga}},
  \bibinfo{author}{\bibfnamefont{R.~R.~R.} \bibnamefont{Reis}},
  \bibinfo{author}{\bibfnamefont{M.~O.} \bibnamefont{Calv\~ao}},
  \bibnamefont{and} \bibinfo{author}{\bibfnamefont{B.~L.} \bibnamefont{Lago}},
  \bibinfo{journal}{JCAP} \textbf{\bibinfo{volume}{03}}, \bibinfo{pages}{027}
  (\bibinfo{year}{2012}), \eprint{1203.3213}.

\bibitem[{\citenamefont{Roman-Garza et~al.}(2019)\citenamefont{Roman-Garza,
  Verdugo, Magana, and Motta}}]{Roman-Garza:2018cxf}
\bibinfo{author}{\bibfnamefont{J.}~\bibnamefont{Roman-Garza}},
  \bibinfo{author}{\bibfnamefont{T.}~\bibnamefont{Verdugo}},
  \bibinfo{author}{\bibfnamefont{J.}~\bibnamefont{Magana}}, \bibnamefont{and}
  \bibinfo{author}{\bibfnamefont{V.}~\bibnamefont{Motta}},
  \bibinfo{journal}{Eur. Phys. J. C} \textbf{\bibinfo{volume}{79}},
  \bibinfo{pages}{890} (\bibinfo{year}{2019}), \eprint{1806.03538}.

\bibitem[{\citenamefont{Chevallier and Polarski}(2001)}]{Chevallier:2000qy}
\bibinfo{author}{\bibfnamefont{M.}~\bibnamefont{Chevallier}} \bibnamefont{and}
  \bibinfo{author}{\bibfnamefont{D.}~\bibnamefont{Polarski}},
  \bibinfo{journal}{Int. J. Mod. Phys.} \textbf{\bibinfo{volume}{D10}},
  \bibinfo{pages}{213} (\bibinfo{year}{2001}), \eprint{gr-qc/0009008}.

\bibitem[{\citenamefont{Linder}(2003)}]{Linder:2002et}
\bibinfo{author}{\bibfnamefont{E.~V.} \bibnamefont{Linder}},
  \bibinfo{journal}{Phys. Rev. Lett.} \textbf{\bibinfo{volume}{90}},
  \bibinfo{pages}{091301} (\bibinfo{year}{2003}), \eprint{astro-ph/0208512}.

\bibitem[{\citenamefont{Yang et~al.}(2021)\citenamefont{Yang, Di~Valentino,
  Pan, Wu, and Lu}}]{Yang:2021flj}
\bibinfo{author}{\bibfnamefont{W.}~\bibnamefont{Yang}},
  \bibinfo{author}{\bibfnamefont{E.}~\bibnamefont{Di~Valentino}},
  \bibinfo{author}{\bibfnamefont{S.}~\bibnamefont{Pan}},
  \bibinfo{author}{\bibfnamefont{Y.}~\bibnamefont{Wu}}, \bibnamefont{and}
  \bibinfo{author}{\bibfnamefont{J.}~\bibnamefont{Lu}}, \bibinfo{journal}{Mon.
  Not. Roy. Astron. Soc.} \textbf{\bibinfo{volume}{501}}, \bibinfo{pages}{5845}
  (\bibinfo{year}{2021}), \eprint{2101.02168}.

\bibitem[{\citenamefont{Russo}(2004)}]{Russo:2004ym}
\bibinfo{author}{\bibfnamefont{J.~G.} \bibnamefont{Russo}},
  \bibinfo{journal}{Phys. Lett. B} \textbf{\bibinfo{volume}{600}},
  \bibinfo{pages}{185} (\bibinfo{year}{2004}), \eprint{hep-th/0403010}.

\bibitem[{\citenamefont{Bilic et~al.}(2005)\citenamefont{Bilic, Tupper, and
  Viollier}}]{Bilic:2005sp}
\bibinfo{author}{\bibfnamefont{N.}~\bibnamefont{Bilic}},
  \bibinfo{author}{\bibfnamefont{G.~B.} \bibnamefont{Tupper}},
  \bibnamefont{and} \bibinfo{author}{\bibfnamefont{R.~D.}
  \bibnamefont{Viollier}}, \bibinfo{journal}{JCAP}
  \textbf{\bibinfo{volume}{10}}, \bibinfo{pages}{003} (\bibinfo{year}{2005}),
  \eprint{astro-ph/0503428}.

\bibitem[{\citenamefont{Carvalho et~al.}(2006)\citenamefont{Carvalho, Alcaniz,
  Lima, and Silva}}]{Carvalho:2006fy}
\bibinfo{author}{\bibfnamefont{F.}~\bibnamefont{Carvalho}},
  \bibinfo{author}{\bibfnamefont{J.~S.} \bibnamefont{Alcaniz}},
  \bibinfo{author}{\bibfnamefont{J.}~\bibnamefont{Lima}}, \bibnamefont{and}
  \bibinfo{author}{\bibfnamefont{R.}~\bibnamefont{Silva}},
  \bibinfo{journal}{Phys. Rev. Lett.} \textbf{\bibinfo{volume}{97}},
  \bibinfo{pages}{081301} (\bibinfo{year}{2006}), \eprint{astro-ph/0608439}.

\bibitem[{\citenamefont{Srivastava}(2007)}]{Srivastava:2006xq}
\bibinfo{author}{\bibfnamefont{S.}~\bibnamefont{Srivastava}},
  \bibinfo{journal}{Phys. Lett. B} \textbf{\bibinfo{volume}{648}},
  \bibinfo{pages}{119} (\bibinfo{year}{2007}), \eprint{astro-ph/0603601}.

\bibitem[{\citenamefont{Wu et~al.}(2008)\citenamefont{Wu, Santos, Vo, and
  Wang}}]{Wu:2008sc}
\bibinfo{author}{\bibfnamefont{Q.}~\bibnamefont{Wu}},
  \bibinfo{author}{\bibfnamefont{N.}~\bibnamefont{Santos}},
  \bibinfo{author}{\bibfnamefont{P.}~\bibnamefont{Vo}}, \bibnamefont{and}
  \bibinfo{author}{\bibfnamefont{A.}~\bibnamefont{Wang}},
  \bibinfo{journal}{JCAP} \textbf{\bibinfo{volume}{09}}, \bibinfo{pages}{004}
  (\bibinfo{year}{2008}), \eprint{0804.0620}.

\bibitem[{\citenamefont{Bento et~al.}(2008)\citenamefont{Bento, Felipe, and
  Santos}}]{Bento:2008yx}
\bibinfo{author}{\bibfnamefont{M.}~\bibnamefont{Bento}},
  \bibinfo{author}{\bibfnamefont{R.}~\bibnamefont{Felipe}}, \bibnamefont{and}
  \bibinfo{author}{\bibfnamefont{N.}~\bibnamefont{Santos}},
  \bibinfo{journal}{Phys. Rev. D} \textbf{\bibinfo{volume}{77}},
  \bibinfo{pages}{123512} (\bibinfo{year}{2008}), \eprint{0801.3450}.

\bibitem[{\citenamefont{Fabris et~al.}(2010)\citenamefont{Fabris, Fraga,
  Pinto-Neto, and Zimdahl}}]{Fabris:2009mn}
\bibinfo{author}{\bibfnamefont{J.~C.} \bibnamefont{Fabris}},
  \bibinfo{author}{\bibfnamefont{B.}~\bibnamefont{Fraga}},
  \bibinfo{author}{\bibfnamefont{N.}~\bibnamefont{Pinto-Neto}},
  \bibnamefont{and} \bibinfo{author}{\bibfnamefont{W.}~\bibnamefont{Zimdahl}},
  \bibinfo{journal}{JCAP} \textbf{\bibinfo{volume}{04}}, \bibinfo{pages}{008}
  (\bibinfo{year}{2010}), \eprint{0910.3246}.

\bibitem[{\citenamefont{Bose and Majumdar}(2011)}]{Bose:2010gc}
\bibinfo{author}{\bibfnamefont{N.}~\bibnamefont{Bose}} \bibnamefont{and}
  \bibinfo{author}{\bibfnamefont{A.}~\bibnamefont{Majumdar}},
  \bibinfo{journal}{Mon. Not. Roy. Astron. Soc.}
  \textbf{\bibinfo{volume}{418}}, \bibinfo{pages}{L45} (\bibinfo{year}{2011}),
  \eprint{1010.5071}.

\bibitem[{\citenamefont{Vargas et~al.}(2012)\citenamefont{Vargas,
  Hipolito-Ricaldi, and Zimdahl}}]{Vargas:2011sz}
\bibinfo{author}{\bibfnamefont{C.~Z.} \bibnamefont{Vargas}},
  \bibinfo{author}{\bibfnamefont{W.~S.} \bibnamefont{Hipolito-Ricaldi}},
  \bibnamefont{and} \bibinfo{author}{\bibfnamefont{W.}~\bibnamefont{Zimdahl}},
  \bibinfo{journal}{JCAP} \textbf{\bibinfo{volume}{04}}, \bibinfo{pages}{032}
  (\bibinfo{year}{2012}), \eprint{1112.5337}.

\bibitem[{\citenamefont{Chen et~al.}(2014)\citenamefont{Chen, Gong, Saridakis,
  and Gong}}]{Chen:2011cy}
\bibinfo{author}{\bibfnamefont{X.-m.} \bibnamefont{Chen}},
  \bibinfo{author}{\bibfnamefont{Y.}~\bibnamefont{Gong}},
  \bibinfo{author}{\bibfnamefont{E.~N.} \bibnamefont{Saridakis}},
  \bibnamefont{and} \bibinfo{author}{\bibfnamefont{Y.}~\bibnamefont{Gong}},
  \bibinfo{journal}{Int. J. Theor. Phys.} \textbf{\bibinfo{volume}{53}},
  \bibinfo{pages}{469} (\bibinfo{year}{2014}), \eprint{1111.6743}.

\bibitem[{\citenamefont{Qi et~al.}(2016)\citenamefont{Qi, Yang, Zhang, and
  Liu}}]{Qi:2014yxa}
\bibinfo{author}{\bibfnamefont{J.-Z.} \bibnamefont{Qi}},
  \bibinfo{author}{\bibfnamefont{R.-J.} \bibnamefont{Yang}},
  \bibinfo{author}{\bibfnamefont{M.-J.} \bibnamefont{Zhang}}, \bibnamefont{and}
  \bibinfo{author}{\bibfnamefont{W.-B.} \bibnamefont{Liu}},
  \bibinfo{journal}{Res. Astron. Astrophys.} \textbf{\bibinfo{volume}{16}},
  \bibinfo{pages}{022} (\bibinfo{year}{2016}), \eprint{1403.7287}.

\end{thebibliography}
\end{document}